\documentclass[fleqn,usenatbib]{mnras}
\usepackage{newtxtext,newtxmath}
\usepackage[T1]{fontenc}
\usepackage{ae,aecompl,bm,booktabs}
\usepackage{amsmath,color,xspace,multirow,longtable,graphicx,bigints,comment}
\usepackage{tabularx}
\usepackage{pdflscape}
\usepackage{caption}
\usepackage{subcaption}
\usepackage{mismath,xcolor}

\newcommand\textlcsc[1]{\textsc{\MakeLowercase{#1}}}

\newcommand{\cidthree}{CID\_346\xspace}
\newcommand{\xnfour}{X\_N\_44\_64\xspace}
\newcommand{\xnsix}{X\_N\_6\_27\xspace}

\title[The CGM of high-z AGN]{An investigation of the circumgalactic medium around $\mathbf{z\sim2.2}$ AGN with ACA and ALMA}

\author[G. C. Jones et al.]{
G. C. Jones$^{1}$\thanks{E-mail: gareth.jones@physics.ox.ac.uk},
R. Maiolino$^{2,3,4}$, 
S. Carniani$^{5}$,
C. Circosta$^{6,4}$,
Y. Fudamoto$^{7,8}$,
J. Scholtz$^{2,3}$
\\
$^{1}$Department of Physics, University of Oxford, Denys Wilkinson Building, Keble Road, Oxford OX1 3RH, UK\\
$^{2}$Cavendish Laboratory, University of Cambridge, 19 J. J. Thomson Ave., Cambridge CB3 0HE, UK\\
$^{3}$Kavli Institute for Cosmology, University of Cambridge, Madingley Road, Cambridge CB3 0HA, UK\\
$^{4}$Department of Physics \& Astronomy, University College London, Gower Street, London WC1E 6BT, UK\\
$^{5}$Scuola Normale Superiore, Piazza dei Cavalieri 7, I-56126 Pisa, Italy\\
$^{6}$European Space Agency (ESA), European Space Astronomy Centre (ESAC), Camino Bajo del Castillo s/n, 28692 Villanueva de la Ca\~{n}ada, Madrid, Spain\\
$^{7}$Waseda Research Institute for Science and Engineering, Faculty of Science and Engineering, Waseda University, 3-4-1 Okubo, Shinjuku, Tokyo 169-8555, Japan\\
$^{8}$National Astronomical Observatory of Japan, 2-21-1, Osawa, Mitaka, Tokyo, Japan\\
}
\begin{document}
\label{firstpage}
\pagerange{\pageref{firstpage}--\pageref{lastpage}}
\maketitle

\begin{abstract}
While observations of molecular gas at cosmic noon %($z\sim2$)
and beyond have focused on the gas within galaxies (i.e., the interstellar medium; ISM), it is also crucial to study the molecular gas reservoirs surrounding each galaxy (i.e., in the circumgalactic medium; CGM). Recent observations of galaxies and quasars hosts at high redshift ($z>2$)
%through the tracers [CII]158\,$\mu$m and CO
have revealed evidence for cold gaseous halos of scale $r_{\mathrm{CGM}}\sim10$\,kpc, with one discovery of a molecular halo with $r_{\mathrm{CGM}}\sim200$\,kpc and a molecular gas mass one order of magnitude larger than the ISM of the central galaxy. As a follow-up, we present deep ACA and ALMA observations of CO(3-2) from this source and two other quasar host galaxies at $z\sim2.2$. While we find evidence for CO emission on scales of $r\sim10$\,kpc, we do not find evidence for molecular gas on scales larger than $r>20$\,kpc. Therefore, our deep data do not confirm the existence of massive molecular halos on scales of $\sim100$\,kpc for these X-ray selected quasars. As an interesting by-product of our deep observations, we obtain the tentative detection of a negative continuum signal on scales larger than $r>200$\,kpc, which might be tracing the Sunyaev-Zeldovich effect associated with the halo heated by the active galactic nucleus (AGN). If confirmed with deeper data, this could be direct evidence of the preventive AGN feedback process expected by cosmological simulations.
\end{abstract}

\section{Introduction}\label{intro}

The study of molecular gas is crucial to understand how galaxies have evolved throughout cosmic time. This cold gas is the primary fuel for star formation, so its presence or absence strongly influences the final stellar mass of the galaxy. This may be seen in the similar evolution of the star formation rate density and molecular gas density throughout cosmic time (e.g., \citealt{deca19,khus21}).

Gas may be accreted by the galaxy via cold, filamentary flows (e.g., \citealt{dano15,Bennett20}), or via wet (i.e., gas-rich) mergers (e.g., \citealt{riec08,pesc20}). On the other hand, it may be expelled by outflows (e.g., \citealt{Feruglio10,Fluetsch19,Lutz20,robe20}) or consumed in the process of star formation. These processes act on multiple scales, from the immediate surroundings of the active galactic nucleus (AGN; $\lesssim20$\,pc), to the interstellar medium (ISM, $r\lesssim5$\,kpc) of the galaxy, and extending to the circumgalactic medium (CGM, $r\gtrsim10$\,kpc) that surrounds it. The multi-scale nature of the baryon cycle makes it imperative to study the molecular content of both the ISM and CGM.

Using a number of spectral lines as tracers (e.g., HI, CO, [CI], [CII]), the properties of cold gas in the ISM of local and distant galaxies have been thoroughly studied (e.g., \citealt{cari02,walt11,hunt12,lefe20,bouw21,lero21}). These studies have revealed the gas content, kinematics (i.e., rotation, outflows, and merging), and morphology of galaxies from the local to high-redshift Universe, including a vast variety of galaxy types (e.g., dwarfs, SFGs, starbursts, quasar host galaxies). For the molecular gas specifically, while there are ongoing investigations into how to correctly transform line intensities to molecular gas masses (e.g., \citealt{bola13,madd20,vizg22}), there are decades of observations and study of molecular gas on the ISM-scale \citep[see e.g. review by ][]{Tacconi20}.

%This complex scenario of has also been seen in cosmological zoom-in simulations (e.g., \citealt{pall17}), where 
The CGM of galaxies is also well studied, but primarily in its atomic ionized phases through absorption systems along the line of sight of background luminous sources \citep[e.g. ][]{Werk16,Tumlinson17} and by mapping the Lyman-$\alpha$ line. The latter observations have revealed Ly$\alpha$ `nebulae' that extend $\gtrsim100$\,kpc around protoclusters (e.g., \citealt{stei00,cai17,trava20}) and smaller halos of $\gtrsim10$\,kpc around individual galaxies (e.g., \citealt{stei11,Borisova16,lecl17,Arrigoni19,Guo20,wang21,Ginolfi22}). However, the resonant nature of Ly$\alpha$ makes it non-trivial to extract the physical properties of the emitting gas (e.g., \citealt{haye15}).

The cold phase of the CGM has been explored much less extensively. The primary issue is that while millimetre/submillimetre line tracers of the cold gas may also be used to probe the CGM, the low surface brightness of this spatially extended gas makes direct detection difficult, especially at high redshift. Despite this, detections of cold gas halos of radius $\sim10$\,kpc have been reported through ALMA observations of [CII]158\,$\mu$m observations of $z\gtrsim2$ galaxies (\citealt{fuji19,fuji20,Ginolfi20,herr21,Debreuck22}).

The detection of molecular gas transitions in the CGM has been more challenging. The detection of CO transitions has been obtained in the CGM around radio galaxies and radio quasars, both locally \citep[e.g. ][]{Russell17,Russell19} and at high redshift \citep{Emonts16,Li21}. In AGNs and normal galaxies, molecular halos have been found on scales of a few $\sim$10\,kpc (\citealt{Ginolfi17,jone23,scho23}).

However, it has been pointed out that spatially extended CO emission may have been missed and filtered out by interferometric observations with ALMA and other extended millimetre observatories. The Atacama Compact Array (ACA) offers the possibility to potentially recover emission on large scales ($>10''$). Within this context, an initial, relatively short-integration ACA observation of CO(3-2) emission from a $z\sim2.2$ AGN host galaxy resulted in the discovery of a molecular halo of radius $200$\,kpc \citep{cico21}. This scale is comparable to the virial radius of a Milky Way-like galaxy (e.g., \citealt{dehn06}) as well as the largest Ly$\alpha$ nebulae found around protoclusters. The amount of molecular gas in the CGM on such large scales is huge; more than an order of magnitude larger than the gas in the ISM of the central galaxy (as inferred from previous ALMA observations).
 %A CO detection requires the emitting medium to be slightly enriched in carbon.
 Such a large amount of molecular gas in the CGM may require a large number of unresolved satellite galaxies (although these were undetected by ALMA, Subaru, and Spitzer; \citealt{cico21}) or a fossil record of enormous outflows of enriched gas. Since this result was determined using relatively low-S/N data, follow-up observations are required.

We follow these initial ACA observations of CO(3-2) emission from a $z\sim2.2$ AGN host galaxy with deeper ACA observations, allowing us to probe lower-luminosity emission. In addition, we target CO(3-2) in two comparable AGN host galaxies, in order to see if the extended emission is ubiquitous for these class of objects. These data are complemented with new and archival ALMA observations, in order to compare emission at different spatial scales.

The details of these observations are listed in Section \ref{obsdat}, and our analysis of the line and continuum emission is presented in Section \ref{analysis}. The presence or absence of extended emission is discussed in Section \ref{DISC}, and we conclude in Section \ref{conc}. We assume a standard concordance cosmology ($\Omega_{\Lambda}$,$\Omega_m$,h)=(0.7,0.3,0.7) throughout.

\section{Observations and data reduction}\label{obsdat}
For this work, we consider three $z\sim2.2-2.3$ AGN host galaxies that are part of the SUPER-ALMA sample. These are X-ray selected AGNs at $z\sim2.3$ which were observed with ALMA in CO(3-2) emission \citep{circ21} using band 3. As mentioned, one of them (\cidthree) was also observed with the ACA \citep{cico21}. In this paper we use deeper ACA observations for \cidthree, new ACA observations for two other AGN host galaxies in the SUPER sample (\xnfour and \xnsix), and also much deeper ALMA CO(3-2) new observations for the latter two sources.
%deeper ACA observations for cid\_346 and new ACA observations for two other AGN host galaxies in the SUPER sample.

Although we also present the new ALMA data, in this paper we primarily focus on the analysis of the ACA data, while a more thorough analysis of the ALMA data will be published in a later paper.

The data used in this paper originate from the ALMA+ACA project 2021.1.00327.S (PI: R. Maiolino). In the case of \cidthree we also combine our ACA data with the previous, shallower ACA data from project 2019.2.00118.S (PI: V. Mainieri). The details of these observations are listed in Table \ref{obstable}. For \cidthree we do not have new ALMA data, so we use archival ALMA data from the programme 2016.1.00798.S (PI Mainieri).

The ACA data for \cidthree from project 2019.2.00118.S were previously analysed by \citet{cico21},
%while the ALMA dataset 2016.1.00798.S was analysed by \citet{circ21}.
while the previous ALMA data of the same source were presented in \cite{circ21}.
A cursory inspection of the new ALMA data for \xnsix from project 2021.1.00327.S was presented in a related work \citep{jone23}, but the rest of the new dataset is unexplored.

%Project 2019.2.00118.S observed a single object: the AGN host galaxy \cidthree ($z= 2.2198\pm0.0001$; \citealt{circ21}). One SPW was tuned for CO(3-2) emission (1024\,channels of width 1.9531\,MHz), while two other SPWs covered the nearby continuum: one in the same sideband (256\,channels of width 7.8125\,MHz) and one in the other sideband (1024\,channels of width 1.9531\,MHz).
%Project 2021.1.00327.S contains further observations of \cidthree, as well as \xnfour ($z_{CO}=2.245\pm0.001$) and \xnsix ($z_{CO}=2.2640\pm0.0004$; \citealt{circ21}). For each of these observations, one SPW was tuned for CO(3-2) emission (256\,channels of width 7.8125\,MHz), while three other SPWs covered the nearby continuum: one in the same sideband and two in the other sideband (all with 128\,channels of width 15.6250\,MHz).

\begin{table*}
\centering
\begin{tabular}{l|ccccc|cc}
Source      &   Array   &   Project         &   Dates               &   On-Source Time &   $\mathrm{N_{antennae}}$ & Synthesized Beam & Resolution          \\
 & & & & [hr] & & & [kpc] \\ \hline 
\cidthree   &   ALMA    &   2016.1.00798.S  &   2016 Dec 1          &   0.16                &   43                      & $(1.31''\times1.19''),53.42^{\circ}$ & $10$  \\
            &   ACA     &   2019.2.00118.S  &   2020 Mar 6-8        &   3.36                &   9-11                    & $(16.37''\times12.09''),-68.36^{\circ}$ & $116$   \\
            &   ACA     &   2021.1.00327.S  &   2021 Oct 10-Nov 27  &   4.49                &   8-9                     & - & - \\ \hline
\xnfour     &   ALMA    &   2021.1.00327.S  &   2022 Jan 24-27      &   3.46                &   42-44                   & $(2.24''\times1.91''),-76.79^{\circ}$  & $17$    \\
            &   ACA     &   2021.1.00327.S  &   2021 Oct 6-15       &   4.99                &   8                       & $(17.21''\times13.64''),86.34^{\circ}$ & $126$   \\ \hline
\xnsix      &   ALMA    &   2021.1.00327.S  &   2022 Jan 23-24      &   2.77                &   42-44                   & $(2.42''\times1.83''),-83.34^{\circ}$  & $17$  \\
            &   ACA     &   2021.1.00327.S  &   2021 Oct 15-29      &   4.99                &   8-10                    & $(18.88''\times11.61''),-61.06^{\circ}$ & $122$   \\ \hline
\end{tabular}
\caption{Properties of ACA \& ALMA observations analysed in this work. We also note the synthesized beam size of each continuum image (see Section \ref{contsec}) and the approximate physical scale of this beam (i.e., the geometric mean of the FWHM values). For the ACA data of \cidthree, we present the beam and resolution information from the combined dataset of both projects.}
\label{obstable}
\end{table*}

The data for each project were downloaded from the ALMA data archive\footnote{\url{https://almascience.eso.org/aq/}}. We then applied the ALMA staff calibration by running the \textit{scriptforPI.py} script using the appropriate CASA version (\citealt{mcmu07};
%4.7.0 for 2016.1.00798.S,
5.6.1 for 2019.2.00118.S, 6.2.7.1 for 2021.1.00327.S). After further inspection, the data were found to not require additional flagging or re-calibration.

The two ACA observations of \cidthree were combined for this analysis (CASA \textlcsc{concat}). We note that while ALMA data for \xnfour and \xnsix were also taken in project 2016.1.00798.S \citep{circ21}, the corresponding integration times are much shorter ($\sim0.15$\,hr) than our new ALMA observations, so we choose not to merge them and to only use our new, high-sensitivity data (Table \ref{obstable}). In what follows, we will refer to the 12\,m array as ALMA and the 7\,m array as ACA (i.e., no combination of antenna types).

More detailed steps of the data processing beyond calibration are provided separately for the continuum and CO-line analysis, and for each individual source, in the following Section.

%also contains ALMA observations of \xnfour and \xnsix, they are very low sensitivity due to their low on-source time ($\sim0.15$\,hr). Because of this, we proceed only with the higher-sensitivity ALMA observations of project 2021.1.00327.S for these two sources.

%In this work, we will combine these archival and new data to explore the presence and extent of continuum and CO emission using several methods.

\section{Analysis}\label{analysis}

\subsection{Image-plane continuum analysis}\label{contsec}

\begin{figure*}
\centering
\includegraphics[width=\textwidth]{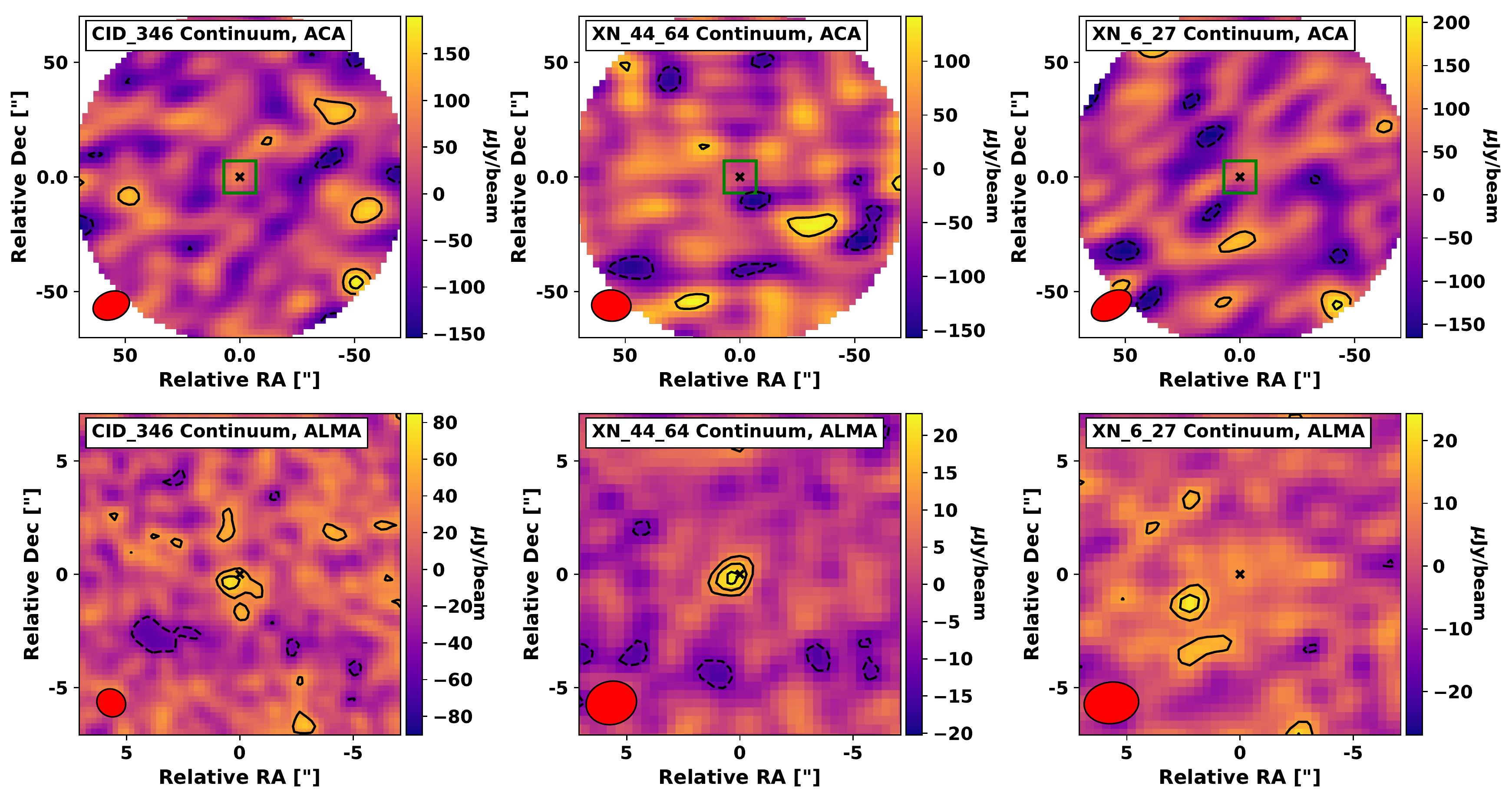}
\caption{Continuum images ($\lambda_{\mathrm{rest}}\sim870\,\mu m$) for each galaxy. Contours are displayed at significance levels of $\pm(2,3,4\ldots)\times\sigma$, where $1\sigma$ is the RMS noise level of the image. The top row shows ACA data (left to right: $1\sigma=[60,50,60]\,\mu$Jy\,beam$^{-1}$). The lower row shows ALMA data (left to right: $1\sigma=[20,5,7]\,\mu$Jy\,beam$^{-1}$). The field of view for the lower images is shown as a green rectangle in the upper row. Synthesized beams and galaxy positions given by red ellipses and black crosses, respectively. Images are cleaned to $3\sigma$, but primary beam correction has not been performed.}
\label{contimages}
\end{figure*}

As a first step, we create rest-frame FIR continuum images for each source. In order to conservatively exclude line emission, we only include channels with $|v|>1000$\,km\,s$^{-1}$ from $z_{CO}$. Here, $z_{CO}$ for \cidthree and \xnfour are taken from \citet{circ21}, while we adopt the revised value of $z_{CO}$ for \xnsix ($z=2.281$) which is found by our analysis (see Section \ref{ipco}). The CASA task \textit{tclean} is used in multi-frequency synthesis (MFS) mode with natural visibility weighting and a primary beam limit of $20\%$ to create `dirty' images. The RMS noise level ($\sigma$) is determined using maps without primary beam corrections. We then clean down to $3\sigma$, resulting in the images presented in Figure \ref{contimages}. None of the ACA continuum images show significant emission, while two of the galaxies are detected in the ALMA images (see Table \ref{bestfitspec}). Below, we briefly discuss the limits on FIR continuum emission that this implies. 

\subsubsection{\cidthree}
A previous analysis of the ALMA data revealed FIR continuum emission from \cidthree ($149\pm43$\,$\mu$Jy; \citealt{circ21}) using a beam of $\sim1''$. A two-dimensional Gaussian fit to our ALMA continuum map returns a similar value ($133\pm57$\,$\mu$Jy). Note that we use the same data as the previous analysis, so there is no improvement in sensitivity. 
While the CASA two-dimensional fitting routine returns an intrinsic (i.e., deconvolved) source FWHM of $(1.4\pm0.8)''\times(0.9\times0.6)''$, the shape of the emission is likely disrupted by noise, and the true emission is unresolved. Using the geometric mean of the half-widths at half-maximum (HWHMs) of the major and minor axes of the beam as an upper size limit (e.g., \citealt{miet17}), this implies a size of $r\lesssim0.6''\sim5$\,kpc.

Since the relatively high-resolution ALMA observations show compact emission, we should detect the continuum emission as a point source at $\sim2-3\sigma$ in the ACA map. However, it is undetected in the ACA map, suggesting a $3\sigma$ upper limit of $<180$\,$\mu$Jy.
This is in agreement with the ALMA value, so we may state that the only FIR continuum emission detected in this source is compact ($<1''$) with an integrated flux density of $133\pm57$\,$\mu$Jy.

\subsubsection{\xnfour}
This source was not detected in FIR continuum emission with ALMA by \citet{circ21}, implying a $3\sigma$ limit of $<66$\,$\mu$Jy. However, this analysis only used 0.15\,hours of on-source observation time. Our much deeper observations (t$_{\mathrm{on-source}}\sim3.5$\,hours) reach a lower RMS noise level ($5\,\mu$Jy\,beam$^{-1}$), and reveal emission with an integrated flux density of $15.6\pm3.0$\,$\mu$Jy. This FIR continuum emission is unresolved, implying a small size ($\lesssim1.0''\sim8$\,kpc), similar to that of \cidthree. Again, we do not detect FIR continuum emission in the ACA data, but this is due to the higher RMS noise level ($50\,\mu$Jy\,beam$^{-1}$). The resulting $3\sigma$ upper limit on the ACA continuum flux density ($<150$\,$\mu$Jy) is consistent with the ALMA detection.

\subsubsection{\xnsix}
Neither the ACA nor the ALMA data show a significant detection of FIR continuum emission, implying $3\sigma$ upper limits of $<180$\,$\mu$Jy and $<21$\,$\mu$Jy, respectively. These are in agreement with the upper limit of \citet{circ21}: $<66$\,$\mu$Jy. We note that there is a $3\sigma$ feature in the ALMA map, but it is separated from the phase centre by $\sim1$ beamwidth and is not coincident with CO emission (Section \ref{ipco}), so we consider it to be noise.

\begin{figure}
    \centering
    \includegraphics[width=0.5\textwidth]{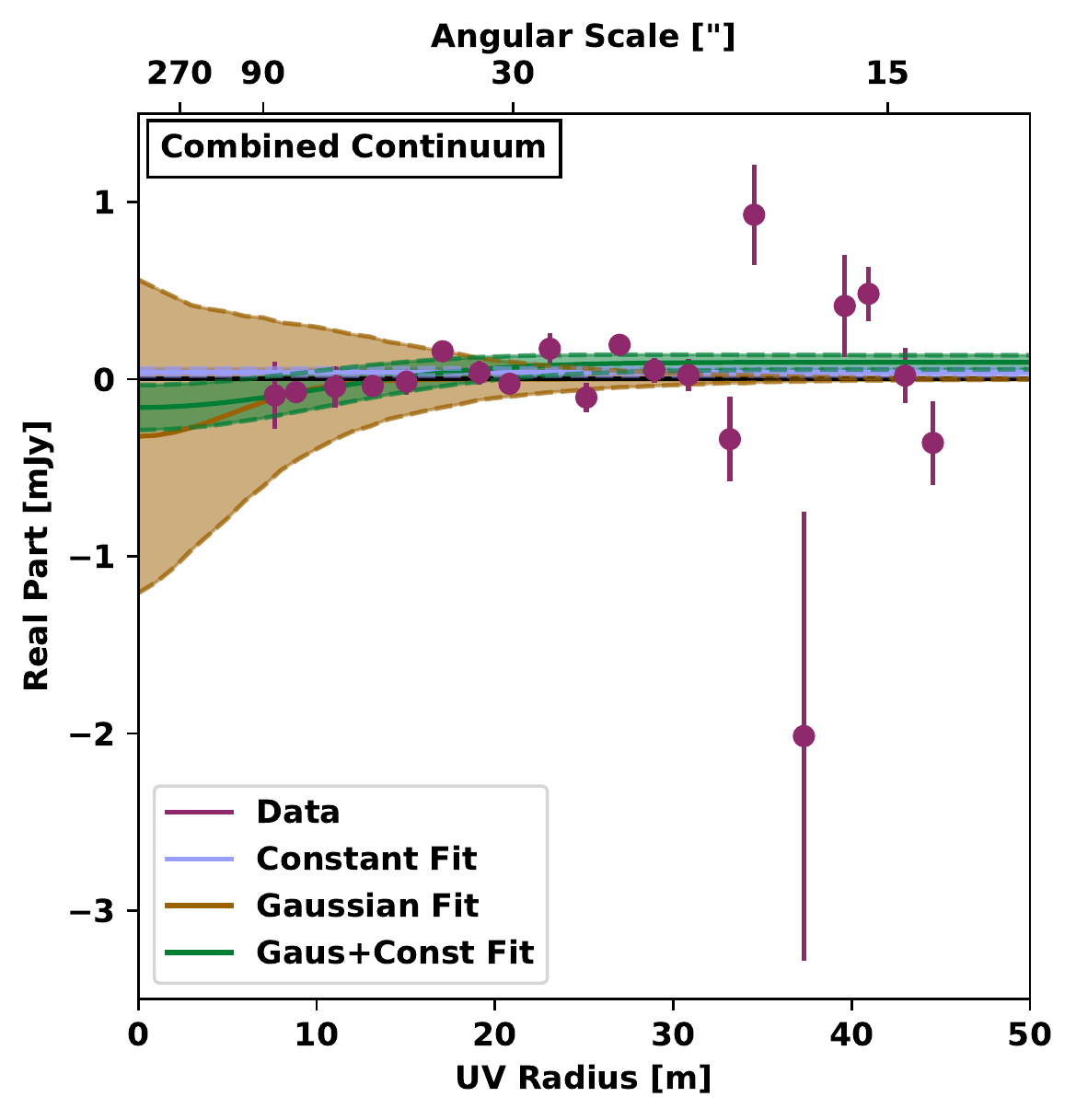}
    \includegraphics[width=0.5\textwidth]{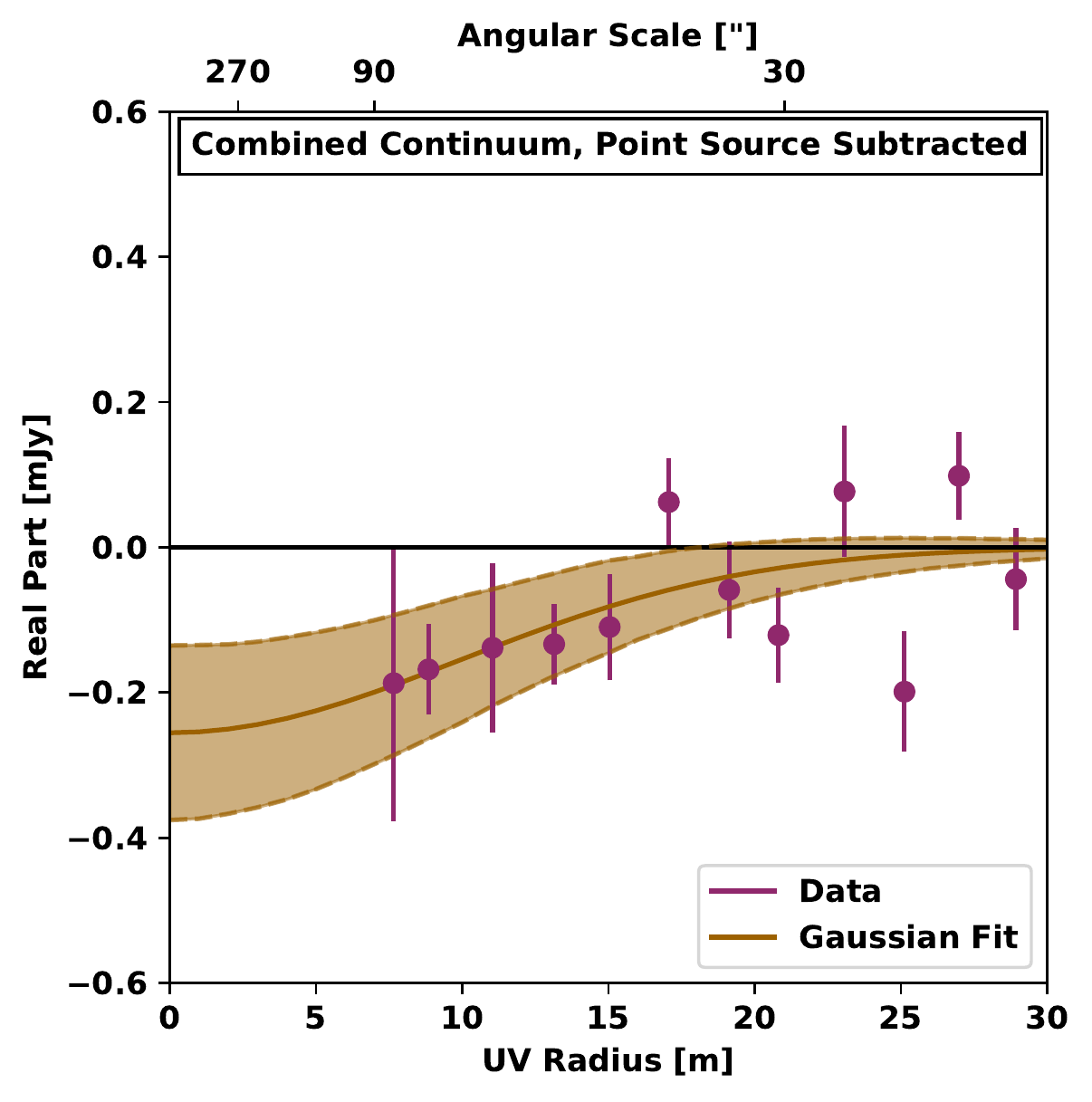}
    \caption{Radial profile of the real part of the stacked observed continuum visibilities from ACA data, including only channels without line emission (maroon points). Top: The results of fitting three models are shown: a single Gaussian (brown line), a constant value (light blue line), and an offset Gaussian (green line). Uncertainties shown by shaded regions ($1\sigma$). Bottom: Results of fitting an offset Gaussian model, subtracting the best-fit offset from the model, and fitting a new Gaussian model. Best-fit parameters and goodness of fit given in Table \ref{vistable}. Note that this is zoomed in with respect to the upper panel.}
    \label{cont_vis}
\end{figure}

\subsection{\textit{uv}-plane continuum analysis}\label{uvcont}
While we do not detect continuum emission in the ACA continuum images of any of the three galaxies, it may be possible that a low-level extended signal is present in the visibilities (i.e., \textit{uv}-data). Here, we examine the stacked continuum visibilities of our ACA data to search for an extended signal. We do not include ALMA data here, as we are interested in the behaviour on large scales (i.e., on short baselines).

To begin, we use the CASA task \textit{split} to separate the line-free channels of each ACA measurement set, while performing time averaging of 30\,seconds (given the short baselines of ACA this time averaging does not affect the angular resolution). These visibilities are then combined into one file (i.e., stacked) and converted into a text file with columns of \textit{u}, \textit{v}, Real(V), Imaginary(V), and the associated weight using the \textlcsc{export\_uvtable} task of the uvplot package \citep{tazz17}. This same package is then used to bin the real part of the visibilities with a given \textit{uv}-bin size (here 2\,m). Note that this procedure is identical to other `stacking' analyses (e.g., \citealt{fuji19}). 

The resulting plot of the real part of the visibilities is shown in Figure \ref{cont_vis} (magenta points). We fit the visibility distribution in \textit{uv}-space with three 1-D models: a constant value, Gaussian, and Gaussian with constant offset. The form of these fits informs us of the spatial distribution of the signal in image space: a constant positive value represents a point source, a Gaussian represents a resolved source, and a Gaussian with constant offset represents the combination of a point source and resolved component. 

The best-fit parameters and goodness of fit values are presented in Table \ref{vistable}, along with the associated $\chi^2$ and reduced $\chi^2$ (hereafter $\chi_{red}^2$). The offset Gaussian fit returns the best $\chi^2$ value as well as the best $\chi_{red}^2$ value, suggesting that the more complex model is not overfitting the data. Both the constant and offset Gaussian models have best-fit constant amplitudes, which represent compact emission, that are in agreement (i.e., within $3\sigma$) with no clear signal. This implies that even a combination of all visibilities does not show a robust continuum signal for the stacked galaxies. This lack of significant compact signal agrees with the weak combined continuum signal as derived in the image-plane analyses of Section \ref{contsec} for the relatively high-resolution observations of ALMA ($0.15\pm0.06$\,mJy; Table \ref{bestfitspec}), and for the ACA observations ($<0.51$\,mJy).

However, one interesting aspect is that both Gaussian models (i.e., `Gaussian Fit` and `Constant+Gaussian Fit') feature negative amplitudes (see Table \ref{vistable}). To show this more clearly, we subtract the best-fit offset from the Constant+Gaussian model and fit these residuals with a Gaussian model (lower panel of Figure \ref{cont_vis}). The negative signal is visible on large angular scales (i.e., $\gtrsim60''$, or $\gtrsim500$\,kpc). This is also true for the simpler `Gaussian' fit, although with a larger uncertainty.

Even though the detection is marginal (i.e., $\sim 2\sigma$ using the fitting uncertainty for the Gaussian amplitude), this is the sort of signal expected to be produced by the Sunyaev-Zel’dovich effect and resulting from the heating of the CGM resulting from the action of AGN feedback (e.g., \citealt{brow19,Lacy19}). This process is expected to be a key phase in galaxy evolution, as the AGN heating of the CGM should prevent cold accretion and therefore result in the quenching of star formation in the galaxy, as a consequence of starvation.  It is particularly interesting to note that the observed signal ($\sim-0.2$\,mJy) is exactly at the level expected from cosmological simulations for this phenomenon \citep[see figure 16 in ][]{brow19}. If confirmed with higher signal-to-noise, this would be an unambiguous confirmation of the preventive, delayed AGN feedback at work at cosmic noon, as expected by models.
 However, our tentative detection should be confirmed with deeper ACA data, or single-dish observations that could fill in the low-\textit{uv} space (e.g., AtLAST; \citealt{klaa20}).

\begin{table*}
\centering
\begin{tabular}{cc|cccc|cc}
                   &                       & Constant      & Gaussian        & Gaussian        & Gaussian        &          & \\
Data               & Model                 & Amplitude     & Amplitude       & $c$             & $c$             & $\chi^2$ & $\chi_{red}^2$  \\ 
                   &                       &[mJy]           &[mJy]           &[m]               &[$''$]               \\ \hline \hline
Stacked Continuum  & Constant Fit          & $0.04\pm0.02$ & $\times$       & $\times$        & $\times$        & 51.05     & 2.69\\
                   & Gaussian Fit          & $\times$      & $-0.3\pm0.9$  & $5.18\pm4.67$   & $122^{+1115}_{-58}$        & 52.89     & 2.94\\
                   & Constant+Gaussian Fit & $0.09\pm0.04$ & $-0.26\pm0.12$ & $9.95\pm4.2$    & $63^{+46}_{-19}$        & 42.84     & 2.52\\ \hline
\cidthree CO(3-2)  & Constant Fit          & $3.51\pm0.48$ & $\times$        & $\times$        & $\times$        & 19.31    & 1.02\\
                     & Gaussian Fit          & $\times$      & $3.51\pm0.76$   & $>10^4$         & $<0.06$        & 19.31    & 1.07\\
                     & Constant+Gaussian Fit & $3.25\pm0.53$ & $34\pm182$      & $3.84\pm4.01$   & $150^{+\infty}_{-77}$        & 16.96    & 1.0\\ \hline
\xnfour CO(3-2)      & Constant Fit          & $1.23\pm0.43$ & $\times$        & $\times$        & $\times$        & 24.41    & 1.74\\
                     & Gaussian Fit          & $\times$      & $8.6\pm6.5$     & $6.62\pm2.23$   & $88^{+45}_{-22}$        & 20.23    & 1.56\\
                     & Constant+Gaussian Fit & $0.67\pm0.57$ & $11.4\pm17.0$   & $5.29\pm2.86$   & $110^{+129}_{-39}$        & 18.8     & 1.57\\
\end{tabular}
\caption{Best-fit values for three models applied to the radial profile of the real part of the visibilities in Figure \ref{cont_vis} (stacked continuum visibilities, see Section \ref{uvcont}) and Figure \ref{uvcont_fig} (line visibilities, see Section \ref{uvline}). We also note the goodness of fit using two statistics.}
\label{vistable}
\end{table*}

\begin{figure*}
    \centering
    \includegraphics[width=\textwidth]{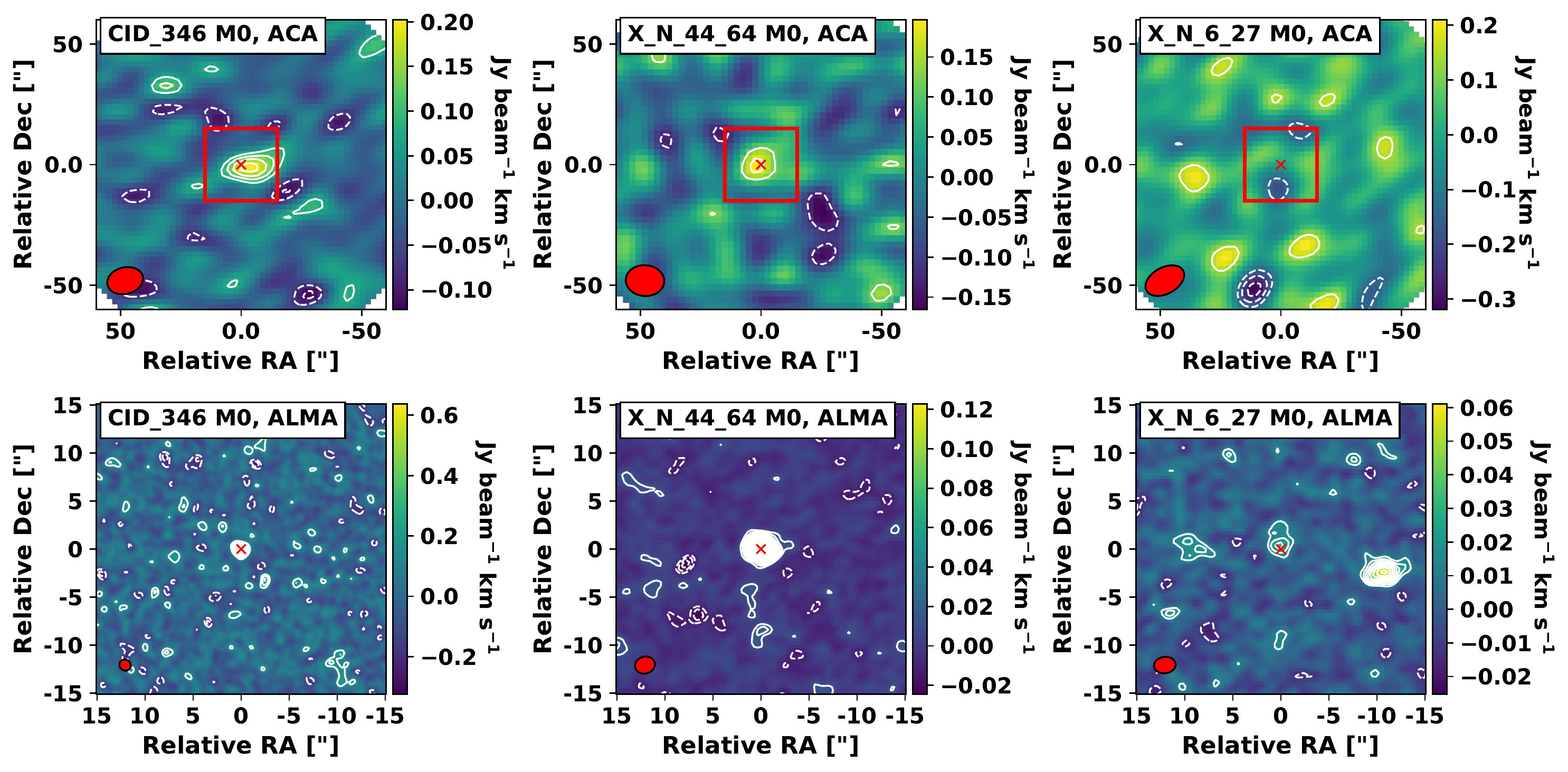}
    \caption{CO(3-2) moment zero maps for each galaxy. Contours are displayed at significance levels of $\pm(2,3,4\ldots)\times\sigma$, where $1\sigma$ is the RMS noise level of the image. The top row shows ACA data (left to right: $1\sigma=[0.04,0.06,0.07]\,$Jy\,beam$^{-1}$\,km\,s$^{-1}$). Lower row shows ALMA data (left to right: $1\sigma=[0.07,0.006,0.007]\,$Jy\,beam$^{-1}$). The field of view for the lower images is shown as a red rectangle in the upper row. Synthesized beam and galaxy position given by red ellipse and cross, respectively. Each map was made by collapsing cleaned data cubes (CASA \textlcsc{immoments}). We choose to present the PB-uncorrected moment zero maps for clarity, as the outskirts of the PB-corrected map feature very high RMS noise levels compared to the central values and are thus obscured by our contours.}
    \label{mom0_6}
\end{figure*}

\subsection{Image-plane CO(3-2) analysis}\label{ipco}

The CO(3-2) emission of all three sources studied in this work have previously been detected in relatively high-resolution (i.e., $\sim1''$) ALMA observations (\citealt{circ21}; although in our data we find a different detection for \xnsix, as explained later in this subsection). Here, we examine new and archival ALMA and ACA data in order to determine the distribution of CO(3-2) emission in each source.

For each dataset, we perform continuum subtraction in the \textit{uv} plane using the CASA task \textit{uvcontsub} to fit a first-order polynomial model to the line-free channels identified in Section \ref{contsec} (i.e., channels width $|v|>1000$\,km\,s$^{-1}$) and subtract this model from the data.

Since the resolution of our ACA observations are an order of magnitude coarser than the previous ALMA observations, the resulting emission should feature a different morphology and higher flux if molecular gas is present on the scales probed by ACA and resolved out by ALMA. To properly explore this putative extended emission, we do not wish to use the previously determined CO properties (i.e., redshift, FWHM, flux density) as priors when characterizing the extended CO emission. Instead, we follow an iterative process. First, we use the CASA task \textit{tclean} with natural visibility weighting and a primary beam limit of $10\%$ to create spectral cubes with channel widths of 15.625\,MHz ($\sim44$\,km\,s$^{-1}$), using only the SPWs containing CO(3-2) emission. The cell size is set to $1/5^{\mathrm{th}}$ of the FWHM of the minor axis of the median synthesized beam. A mean RMS noise level per channel is found ($1\sigma$; CASA \textit{imstat}), and the cube is cleaned down to $3\sigma$. A duplicate of this cube with the primary beam (PB) correction applied is also created. 

A spectrum is extracted from the PB-uncorrected cube using a circular aperture of diameter 10\,px ($\sim20''$ for the ACA data, $\sim2''$ for the ALMA data) centred on the galaxy position, and a list of preliminary channels containing line emission are identified. These channels are collapsed using the CASA task \textit{immoments}, resulting in a moment 0 map. The RMS noise level of this map is measured ($\sigma_{RMS}$), and the $2\sigma_{RMS}$ contour of this emission is used to create a new aperture. A final spectrum is extracted from the PB-corrected data cube using this aperture.

We find that while CO(3-2) emission is detected in all three sources with ALMA, only two sources (\cidthree and \xnfour) show evidence for CO(3-2) emission in the ACA data. More details are given in the following subsections.

\begin{figure*}
    \centering
    \includegraphics[width=\textwidth]{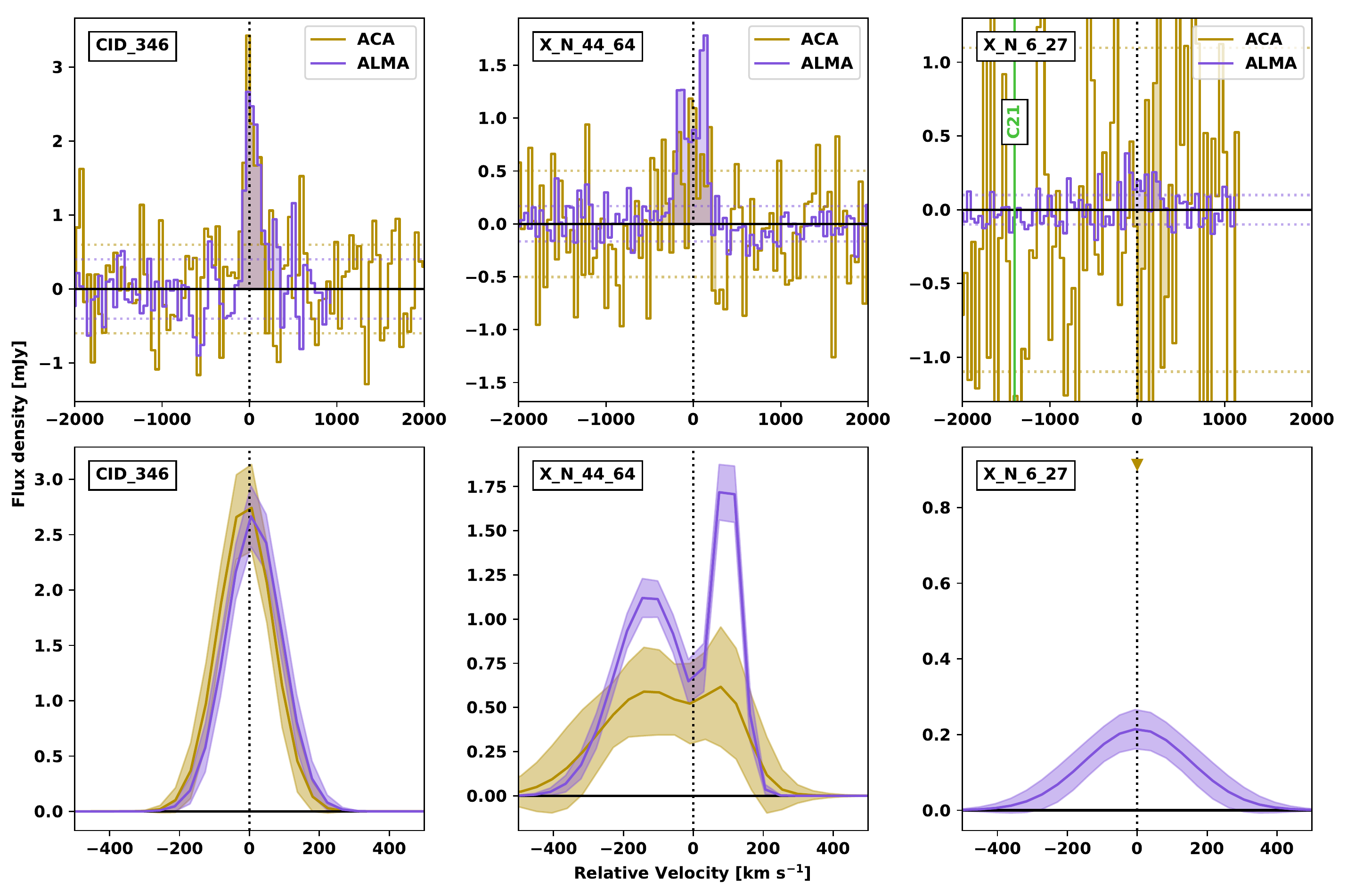}
    \caption{Top row: CO(3-2) spectra for each galaxy, extracted from primary-beam corrected cubes using the $2\sigma$ contours from each moment map (see Figure \ref{mom0_6}). Each panel shows the extracted spectrum from the ACA data (brown) and ALMA data (purple). Channels identified to contain line emission are highlighted. RMS noise level depicted as coloured dotted lines. The redshift of the ALMA detection is shown by a vertical dashed line. Since the ALMA spectrum of \xnfour shows non-Gaussian behaviour, we fit the spectra of this source with two Gaussian components. No significant emission was detected in the ACA data for \xnsix, so we use a 10\,px ($\sim20''$) wide circular aperture instead. The previously determined redshift of \xnsix from \citet{circ21} is shown as a green line in the upper right panel. 
    Lower row: zoomed-in views of the best-fit models for each emission line. The $3\sigma$ upper limit of the average value for the ACA spectrum of \xnsix is shown as a downwards-facing triangle in the lower right plot.}
    \label{spec_6}
\end{figure*}

\begin{table*}
    \centering
    \begin{tabular}{lc|cccc|c}
Source	                &	Array	      			&	Amplitude	    &	Redshift	        &	FWHM	        &	Integrated	        &	S$_{\mathrm{Continuum}}$	\\
	                    &		          			&	[mJy]	        &		                &	[km\,s$^{-1}$]	&	[mJy\,km\,s$^{-1}$]	&	$\mu$Jy	\\ \hline \hline
\cidthree	            &	ALMA	      			&	$2.68\pm0.23$	&	$2.2198\pm0.0001$	&	$187\pm22$	    &	$532\pm84$	        &	$133\pm57$	\\
	                    &	ACA	          			&	$2.81\pm0.40$	&	$2.2196\pm0.0001$	&	$183\pm30$	    &	$548\pm119$	        &	$<180$	\\ \hline
\xnfour$^a$				&	ALMA$_{\mathrm{G1}}$    &	$1.14\pm0.11$   &	$2.2435\pm0.0001$   &	$238\pm33$	    &	$290\pm49$	        &	-	\\
                        &	ALMA$_{\mathrm{G2}}$    &	$1.90\pm0.18$	&	$2.246\pm0.001$     &	$89\pm11$	    &	$179\pm29$	        &	-	\\
                        &	ALMA$_{\mathrm{TOTAL}}$ &	-	            &	$2.2448\pm0.0005$   &	-	            &	$469\pm57$	        &	$15.6\pm3.0$	\\
	                    &	ACA$_{\mathrm{G1}}$     &	$0.59\pm0.26$	&	$2.2435$	        &	$342\pm215$	    &	$216\pm165$	        &	-	\\
	                    &	ACA$_{\mathrm{G2}}$     &	$0.42\pm0.45$	&	$2.246$	            &	$144\pm164$	    &	$64\pm100$	        &	-	\\
	                    &	ACA$_{\mathrm{TOTAL}}$  &	-            	&	$2.2448$            &	-        	    &	$280\pm193$	        &	$<150$	\\ \hline     
\xnsix	    			&	ALMA	      			&	$0.21\pm0.05$	&	$2.2807\pm0.0005$	&	$356\pm99$	    &	$81\pm30$	        &	$<21$	\\
	                    &	ACA$^b$	      			&	$<0.92$	        &	-	                &	-	            &	$<480$	            &	$<180$	\\ \hline
    \end{tabular}
    \caption{Best-fit parameters of one-dimensional Gaussian fits to integrated CO(3-2) spectra and rest-frame FIR continuum flux density. For each source, we present both high- (ALMA) and low-resolution (ACA) results. All upper limits are given as $3\sigma$. $^a$: The ALMA CO(3-2) line profile of \xnfour shows a two-Gaussian nature (see Figure \ref{spec_6}), so we record the best-fit parameters of each peak, as well as the average redshift and combined integrated flux density. Since the ACA CO(3-2) signal for this source is much weaker, we fix the redshifts of each peak to be identical to the best-fit values of the ALMA spectrum and only fit for the amplitudes and linewidths of each peak. $^b$: We find no evidence for CO(3-2) emission in the ACA data of \xnsix, so we only present $3\sigma$ upper limits on its average amplitude and integrated flux density (see Section \ref{xn44co}).}
    \label{bestfitspec}
\end{table*}

\subsubsection{\cidthree}

We use data from the two ACA programs in which \cidthree was observed (see Table \ref{obstable}), selecting only the SPW expected to contain CO(3-2) emission. Using \textit{tclean}, we create a `dirty' image (i.e., without cleaning) with a cellsize of $2.19''$ and a mean RMS noise level per channel of 1.3\,mJy\,beam$^{-1}$. We follow a similar imaging process for the one ALMA program for this source, with a cell size and RMS noise level per channel of $0.22''$ and 0.46\,mJy\,beam$^{-1}$, respectively. 

Moment zero maps of each source are created using the channels identified to contain line emission (Figure \ref{mom0_6}). Using the central $2\sigma_{RMS}$ contour of each map as an aperture we extract a spectrum that reveals a strong CO(3-2) detection (left panels of Figure \ref{spec_6}). Each spectral line is well-fit by a single Gaussian, resulting in low residuals. The properties of each fit are listed in Table \ref{bestfitspec}.

Despite the fact that the ALMA and ACA data feature synthesized beam sizes that differ by an order of magnitude, the CO(3-2) emission recovered by each array is in agreement to within $1\sigma$. This finding is discussed in more detail in Section \ref{extendy}.

\begin{figure}
    \centering
    \includegraphics[width=0.45\textwidth]{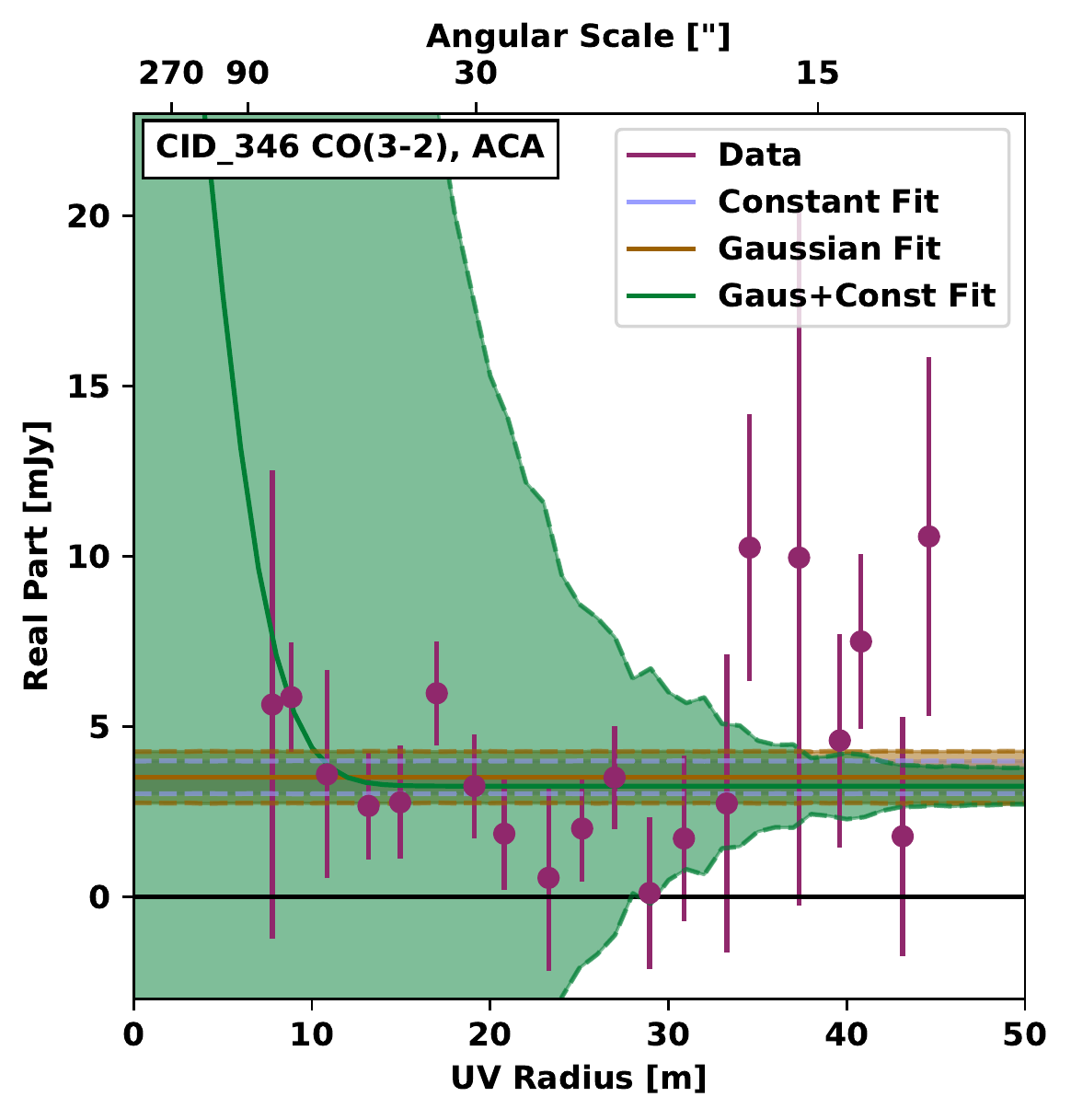}
    \includegraphics[width=0.45\textwidth]{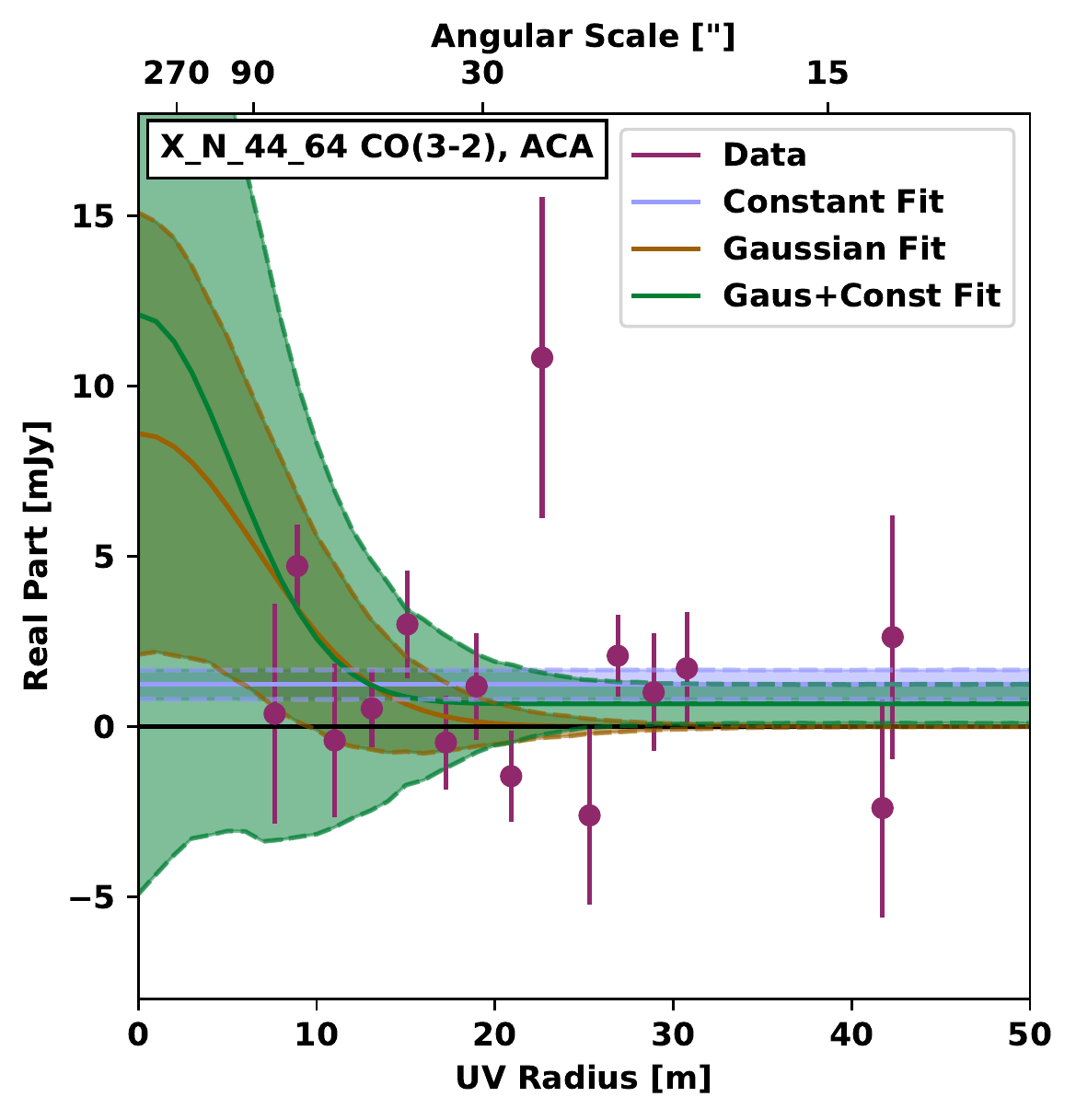}
    \caption{Radial profile of the real part of the CO(3-2) visibilities (maroon points). The results of fitting three models are shown: a single Gaussian (brown line), a constant value (light blue line), and an offset Gaussian (green line). Uncertainties shown by shaded regions ($1\sigma$). Best-fit parameters and goodness of fit values are given in Table \ref{vistable}.}
    \label{uvcont_fig}
\end{figure}

\subsubsection{\xnfour}\label{xn44co}
When imaged, the new ALMA data results in a data cube with a cellsize of $0.35''$ and RMS noise level per channel of $0.16$\,mJy\,beam$^{-1}$. With this high sensitivity, we detect strong CO(3-2) emission (Figure \ref{mom0_6}) with a double-horned profile (central panels of Figure \ref{spec_6}). Since we lack the spatial resolution to decipher the cause of this appearance (e.g., rotation, merging, or outflows), we simply fit two one-dimensional Gaussians to the profile in order to determine the integrated flux density. 

The ACA CO(3-2) cube of \xnfour features an average RMS noise level per channel of 1.4\,mJy\,beam$^{-1}$ and a cellsize of $2.55''$. While the individual channel maps do not show obvious emission, a line is evident using a central circular aperture. The spectrum shows that the line is weak but broad, matching the properties of the previous ALMA observation (\citealt{circ21}). Following the finding that the ALMA CO line is double-peaked, we fit this spectrum with a double-Gaussian model, fixing the redshifts of the two peaks to be identical to those of the ALMA data.

Similarly to \cidthree, the difference between the ALMA and ACA integrated flux densities is not significant ($<2\sigma$), so there is no robust evidence for an extended CO halo from this analysis.

\subsubsection{\xnsix}
Using the data from a single ACA observation program, we create an image of the CO(3-2) emission from this source. The resulting image has a cellsize of $2.16''$ and average RMS noise level per channel of $\sim1.4$\,mJy\,beam$^{-1}$. No significant emission is detected in this data cube (see top right panel of Figure \ref{spec_6}). 

We then image the new ALMA observations of this target, which results in a data cube with a cellsize and RMS noise level per channel of $0.34''$ and 0.19\,mJy\,beam$^{-1}$, respectively. No emission was detected at the originally reported CO-based redshift or the optical line-based redshift of this source ($z=2.2640$ and $z=2.263$; \citealt{circ21}), but there is a $\sim3\sigma$ detection at $z=2.2807$, or $\sim1500$\,km\,s$^{-1}$ from the previous redshift (top-right panel of Figure \ref{spec_6}, where we have shifted the velocity scale to the new redshift), exactly at the phase center (i.e. the location of the galaxy optical counterpart, bottom-right panel of Figure \ref{mom0_6}). Since the RMS noise level per channel of the previous observation was much higher (i.e., $0.51$\,mJy\,beam$^{-1}$), this emission would not be observable without the additional on-source exposure time. 

We also note that there is a strong line detection $\sim11''$ west of the phase centre (bottom-right panel of Fig. \ref{mom0_6}). Since it shows no detectable link to the target galaxy and is bright, discrete, and distant, we conclude that this strong detection at the same redshift of the central detection is likely tracing a physical, gas-rich companion and gives further confidence in the central detection.

Since there is a weak detection of CO(3-2) emission in our ALMA data, we create a moment zero map and extract a spectrum as detailed above. Specifically, in order to illustrate the ACA non-detection, we extract a spectrum from the cleaned, primary beam-corrected data cube using a 10\,px ($\sim20''$) wide circular aperture centred on the expected galaxy position (upper right panel of Figure \ref{spec_6}). In addition, we create a moment zero map using the same channels as in the ALMA moment zero map (Figure \ref{mom0_6} top right panel). Both of these panels show no significant emission.

An upper limit on the ACA CO(3-2) emission of \xnsix can be obtained using $\delta S_{int}=\delta v \sqrt{N_{ch}} RMS_{spec}$, where $\delta v$ is the velocity width of one channel, $N_{ch}$ is the number of channels identified as line emission in the ALMA map, and $RMS_{spec}$ is the RMS noise level of the extracted spectrum\footnote{This equation emerges from propagating the uncertainty on the integrated flux density: $S_{int}\equiv\int S \delta v=\delta v \sum_i S_i$.}. This results in a $3\sigma$ upper limit of $\lesssim480$\,mJy\,km\,s$^{-1}$. Similarly, we may estimate the uncertainty on the average value of the flux density using $\delta \overline{S}=RMS_{spec}/\sqrt{N_{ch}}$, yielding a $3\sigma$ upper limit of $\lesssim0.91$\,mJy.

\subsection{\textit{uv}-plane CO(3-2) analysis}\label{uvline}
The previous image-plane analysis of our ACA CO(3-2) observations show two significant detections. To examine these in a different way, we turn to the line visibilities, as we did for the continuum visibilities in Section \ref{uvcont}. Specifically, we use the CASA task \textit{split} to separate the visibilities corresponding to $\pm$HWHM$_{CO}$ and use \textlcsc{uvplot} to plot the real part of the visibilities as a function of binned \textit{uv}-distance (bins of 2\,m). The resulting plots are shown in Figure \ref{uvcont_fig}, while the best-fit parameters and goodness of fit parameters are listed in Table \ref{vistable}.

The ACA line visibility plot of \cidthree is well fit by a constant value with a significant offset from zero, i.e. unresolved emission. This is strengthened by the fact that the Gaussian fit yields a similar amplitude and very large Gaussian width, which is nearly identical to a constant value. On the other hand, the offset Gaussian model is best fit by a slightly lower constant value and a narrow Gaussian. The fit of this Gaussian component is strongly influenced by the inner three \textit{uv} bins, making the width and amplitude uncertain and fully consistent with a constant value (as shown by the green shaded regions in Figure \ref{uvcont_fig}, which give the uncertainty ranges). We conclude that the visibilities are well represented by a flat model, implying a point source of amplitude $3.51\pm0.48$\,mJy. This is in agreement (i.e., $<2\sigma$ difference) with the image-plane flux density of $2.81\pm0.40$, and also fully consistent with the ALMA flux, which implies a lack of extended emission (see Appendix \ref{appb} for discussion of discussion of a previous ACA CO(3-2) visibility analysis of this source). 

The CO(3-2) emission of \xnfour is less strongly detected than that of \cidthree, so it is not surprising that the best-fit amplitude of a constant model applied to the line visibilities yields a lower amplitude ($1.23\pm0.43$\,mJy). This peak flux is larger than the poorly-fit amplitude of the ACA spectrum ($\sim0.5$\,mJy), but in agreement with the ALMA amplitude ($\sim1-2$\,mJy). This again implies a lack of flux beyond the maximum recoverable scale (MRS) of ALMA. While the two Gaussian models return lower $\chi^2$ and $\chi_{red}^2$ values, their unconstrained Gaussian amplitudes suggest that they are poor fits.

\subsection{CO(3-2) radial profile analysis}\label{HALOMOD}
In the previous subsections, we detailed multiple detections of CO(3-2) using low-resolution observations with the ACA and higher-resolution observations with ALMA. The integrated emission (i.e., moment 0 maps) of these detections do not show obvious extended emissions, with only small deviations from the beam shape. Yet, a recent work \citep{cico21} suggests that this type of source (and \cidthree in particular) may be surrounded by a very large scale ($\sim200$\,kpc) reservoir of molecular gas. To further test the presence of spatially extended emission, we extract radial brightness profiles and test whether these profiles may be explained by an unresolved source, a single resolved component, or a central source with an extended halo.

\subsubsection{Radial profile extraction}

The methods of this analysis is detailed in an associated paper that analyses the ALMA CO(3-2) data of a larger sample of SUPER galaxies \citep{jone23}. In short, we fit a 2-D elliptical Gaussian to the ALMA and ACA CO(3-2) moment zero maps of each source detected in line emission and use the best-fit spatial centroid as the effective centre. We then find the mean value in circular rings of width 1\,px ($\sim2''$ for the ACA data, $\sim0.2''$ for the ALMA data) centred on this position in order to create a radial brightness profile. In parallel, we extract a radial brightness profile from the synthesized beam. The extracted profiles are shown in Figure \ref{PMN_ACA}.

The uncertainty on the mean value is non-trivial to derive, as the noise in each map is correlated on the scale of the synthesized beam. To estimate this effect, we produce 100 maps of pure noise, convolve each with the beam, determine the RMS noise level in each annulus, and take the average value across all maps. The uncertainty is then taken as the greater of this value and the standard deviation of the values in each annulus in the moment zero map. We note that this results in slightly different noise realizations for each run of our code.

\subsubsection{Profile fitting methods}
In order to place constraints on the physical morphology, we first examine whether the emission could be explained by a single unresolved source. In this case, the radial profile of the beam and moment 0 map would be the same. This is tested by finding the $\chi^2$ value between the beam profile and observed emission profile.

Next, we create a circular Gaussian model with a given HWHM, convolve it with the PSF, extract a radial brightness profile, and use the Bayesian inference code PyMultiNest (\citealt{fero09,buch14}) to find the best-fit intrinsic HWHM so that the modelled and observed radial brightness profiles are matched. In this way, we test whether the emission could be explained by a single resolved source. We wish to explore a range of intrinsic widths, so for the ALMA data we fit for log$_{10}($HWHM$_{\mathrm{G1}})$ (where the HWHM is in units of arcseconds) and set the prior to a uniform distribution between [-2.0,1.5], corresponding to angular scales [$0.01'',\sim30''$]. The upper bound of this prior is motivated by the MRS of these observations, which varies between $\sim10''-20''$. Since the ACA data may include much more extended emission, we expand the prior limits to be [-2.0,2.0], corresponding to angular scales [$0.01'',100''$]. Again, this is motivated by the larger MRS of the ACA data ($\sim70''-90''$).

In order to test the existence of a halo around this central source, we add an additional 2-D Gaussian component to the previous model, resulting in three variables: the widths of the two Gaussians (HWHM$_1$, HWHM$_2$) and the relative peak intensity f$_{12}$. For the ALMA data, the prior distributions of the two log$_{10}$(HWHM) variables are set to uniform distributions: [-2,0] (or [$0.01'',1''$]) for log$_{10}($HWHM$_{\mathrm{G1}})$ and [0,1.5] (or [$1'',\sim30''$]) for log$_{10}($HWHM$_{\mathrm{G2}})$. We adopt a uniform distribution between [-3.5,0) for $\mathrm{log_{10}(f_{12})}$. In the case of the ACA data, the HWHM priors are shifted slightly to [-2,0.5] and [0.5,2.0], or [$0.01'',\sim3''$] and [$3'',\sim100''$], respectively.

\begin{figure*}
\includegraphics[width=11.8cm]{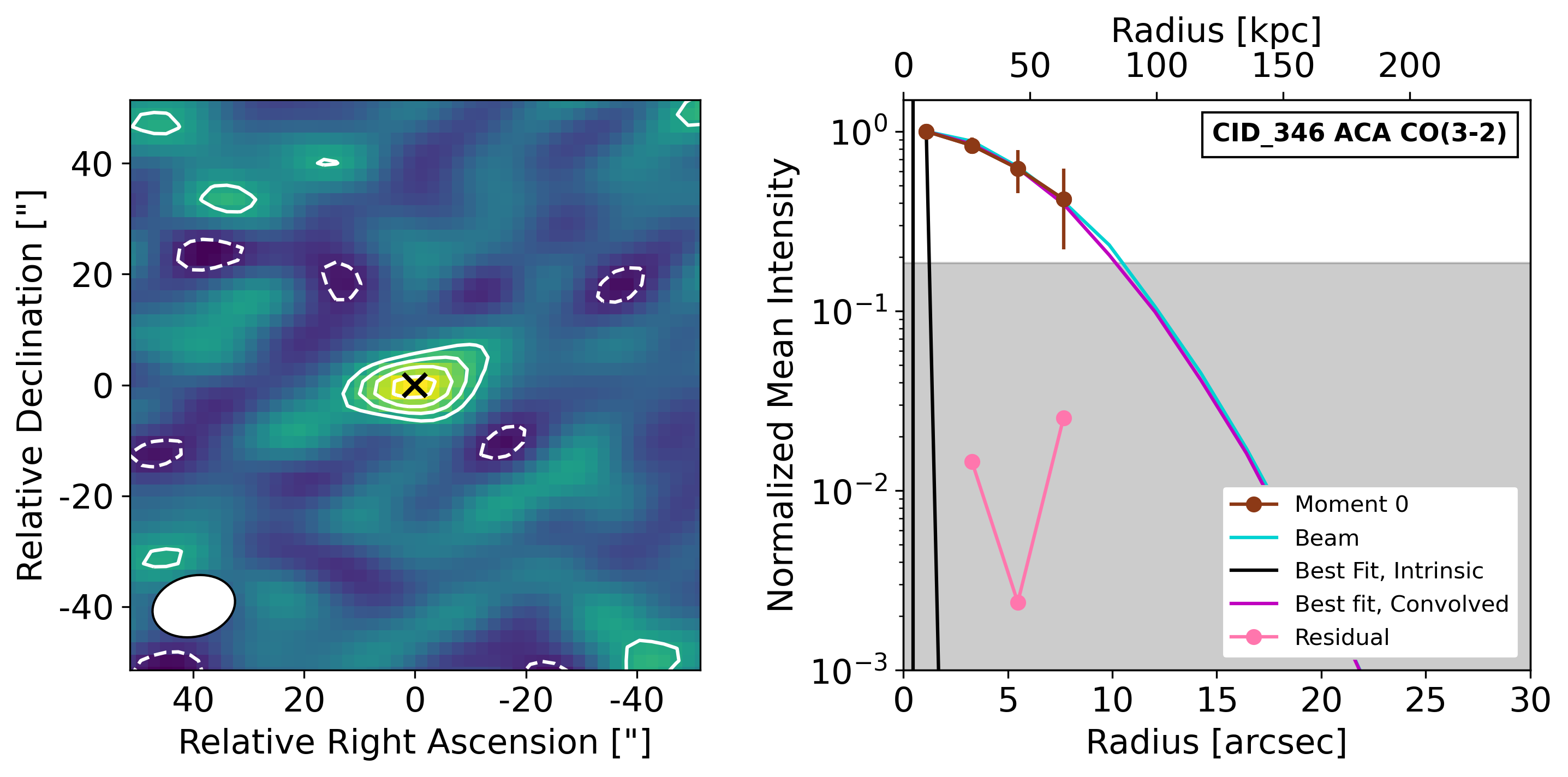}
\includegraphics[trim=11.6cm 0 0 0, clip,width=5.9cm]{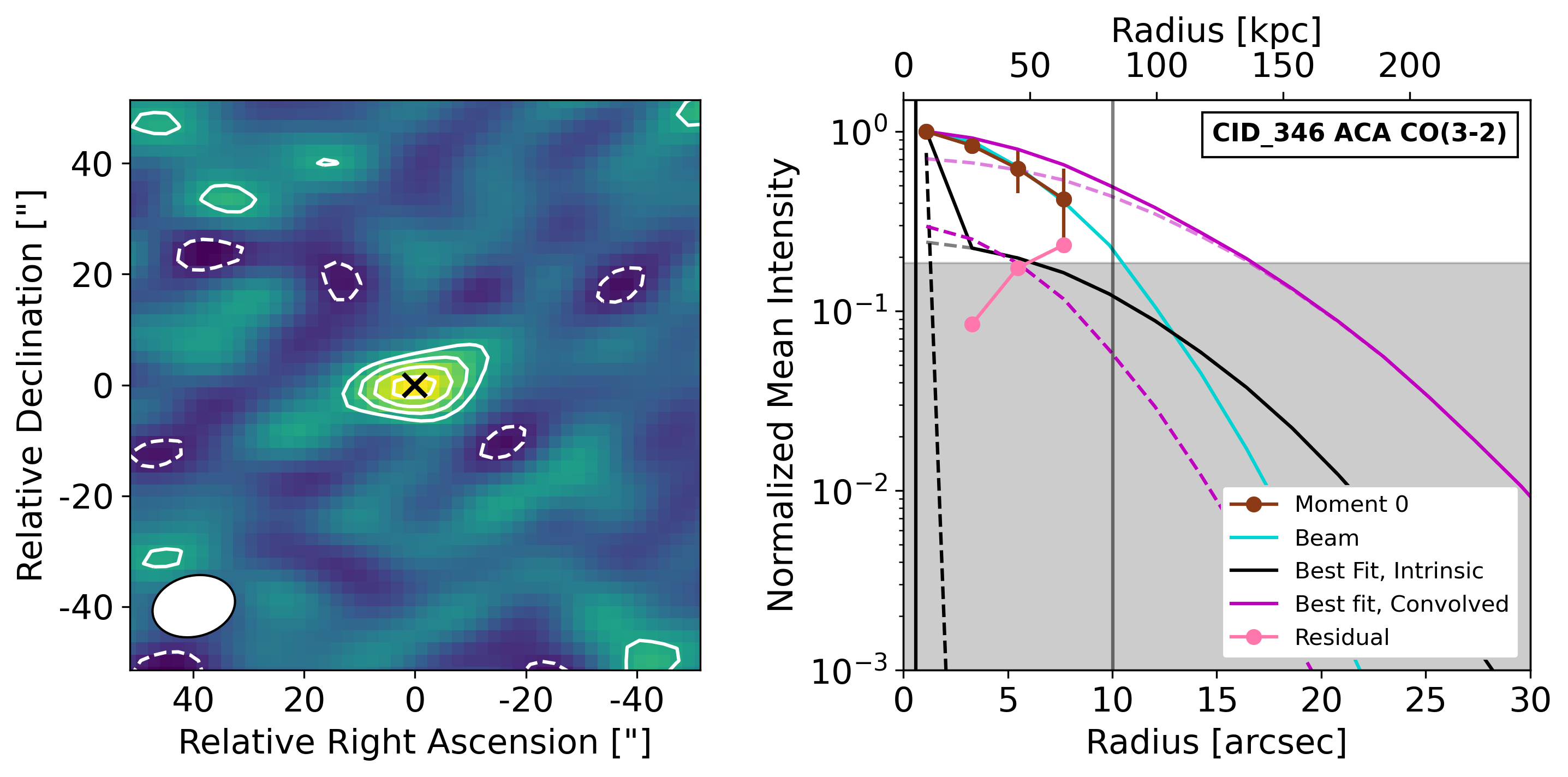}
\includegraphics[width=11.8cm]{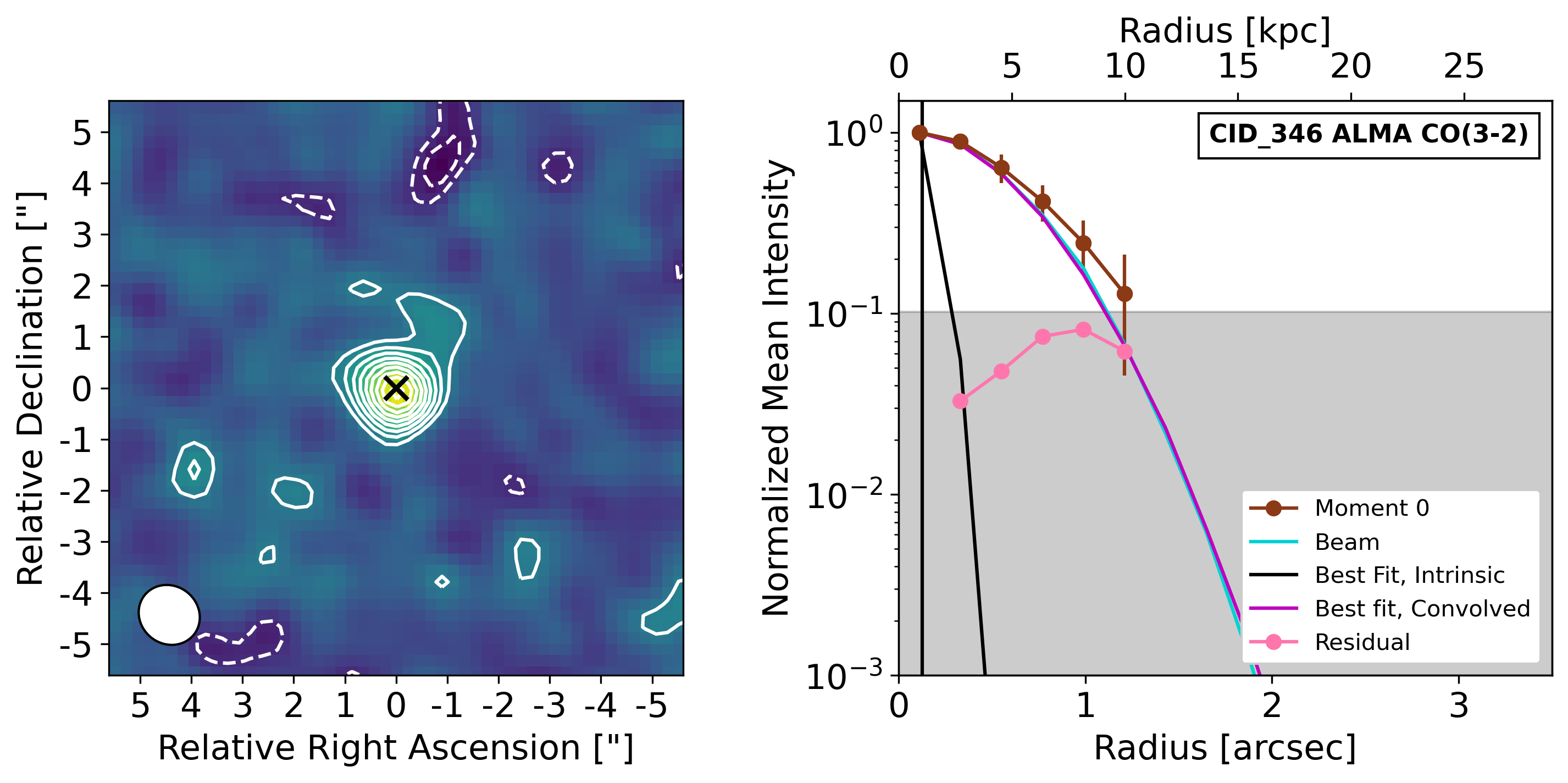}
\includegraphics[trim=11.6cm 0 0 0, clip,width=5.9cm]{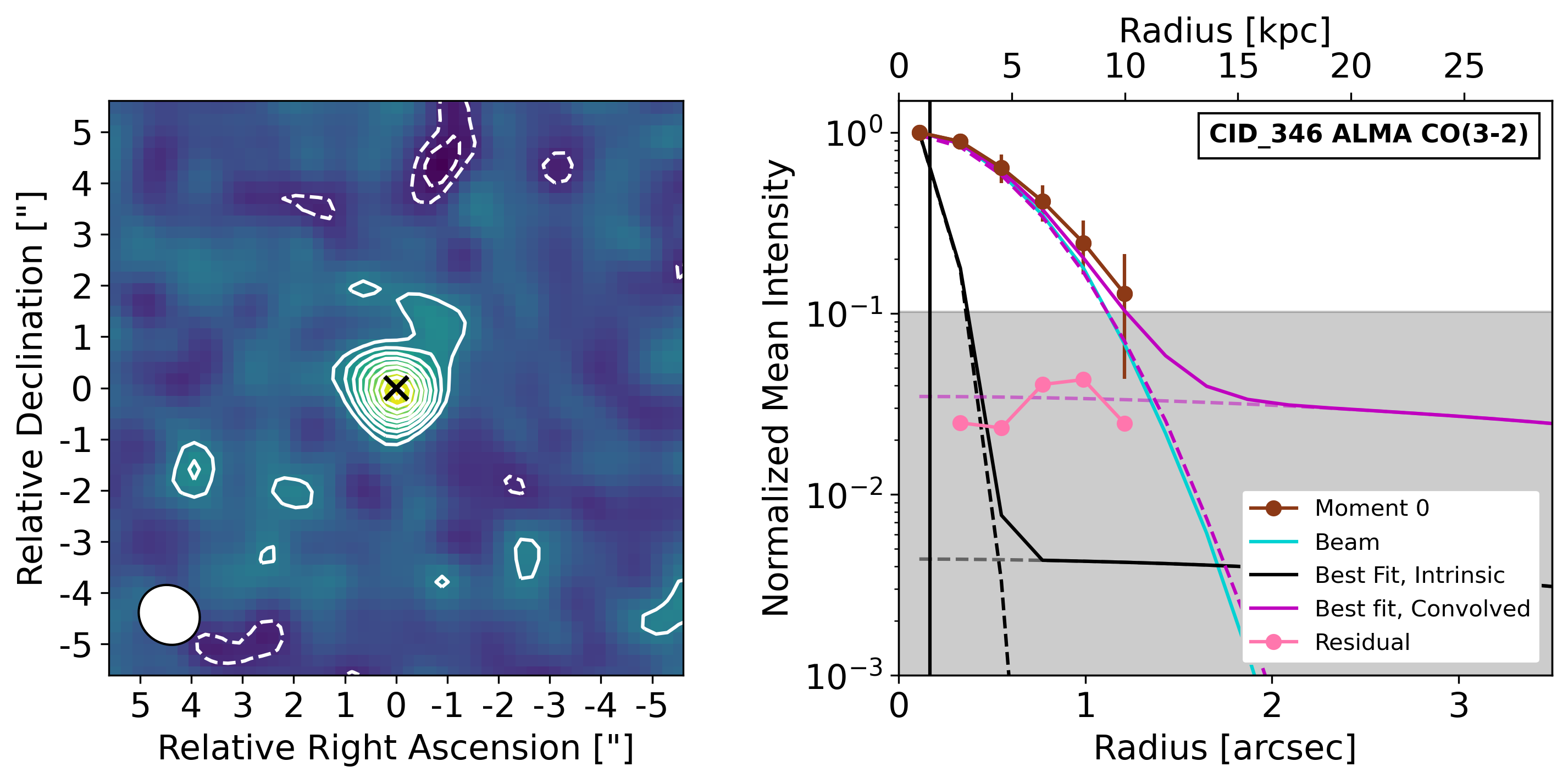}
\includegraphics[width=11.8cm]{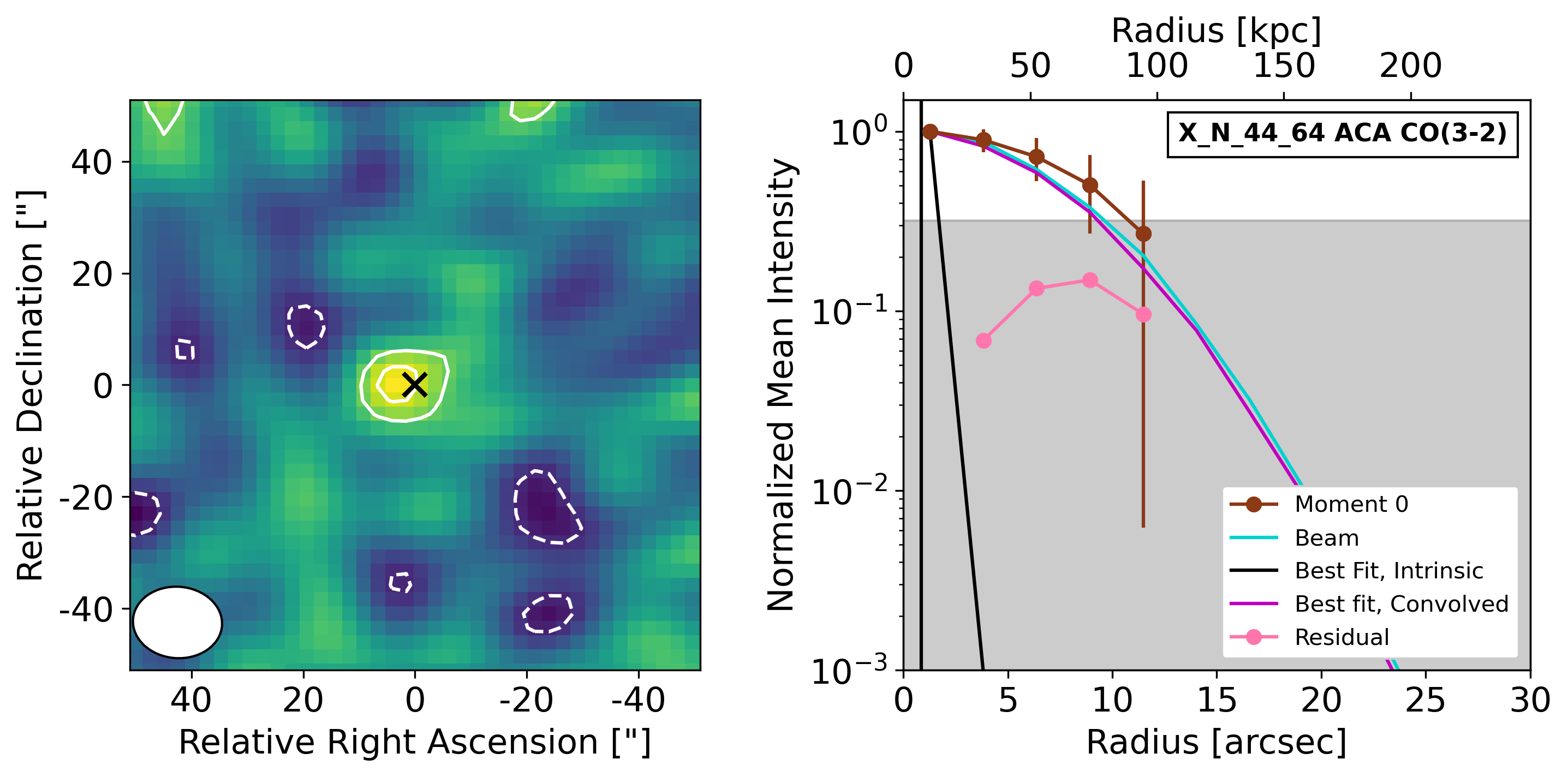}
\includegraphics[trim=11.6cm 0 0 0, clip,width=5.9cm]{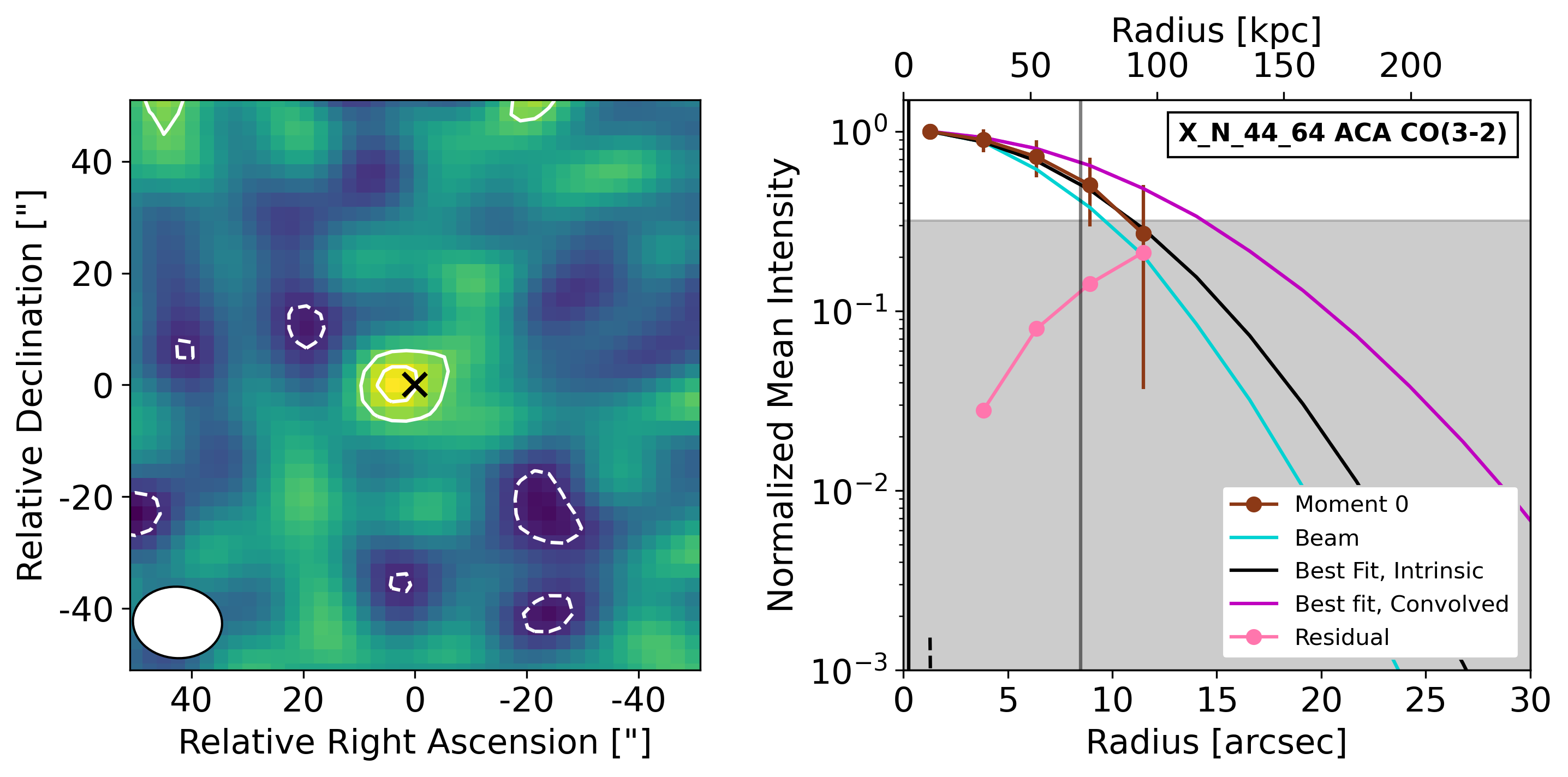}
\caption{Results of fitting a two-dimensional Gaussian model to the moment 0 maps of the \cidthree (top two rows), \xnfour (central 2 rows), and \xnsix (bottom row). Each left panel shows the zoomed in moment zero map of Figure \ref{mom0_6}, while the central and right columns show a single- and two-component model fit, respectively. Brown and cyan lines show the normalized mean radial brightness profiles of the moment 0 map and beam, while the shaded region depicts the RMS noise level of the moment 0 map. We include the best-fit model radial brightness profile (magenta) and intrinsic halo profile (black solid curves). The intrinsic HWHM values are depicted by vertical black line. Each profile is normalized to its maximum value.}
\label{PMN_ACA}
\end{figure*}
\begin{figure*}\ContinuedFloat
\includegraphics[width=11.8cm]{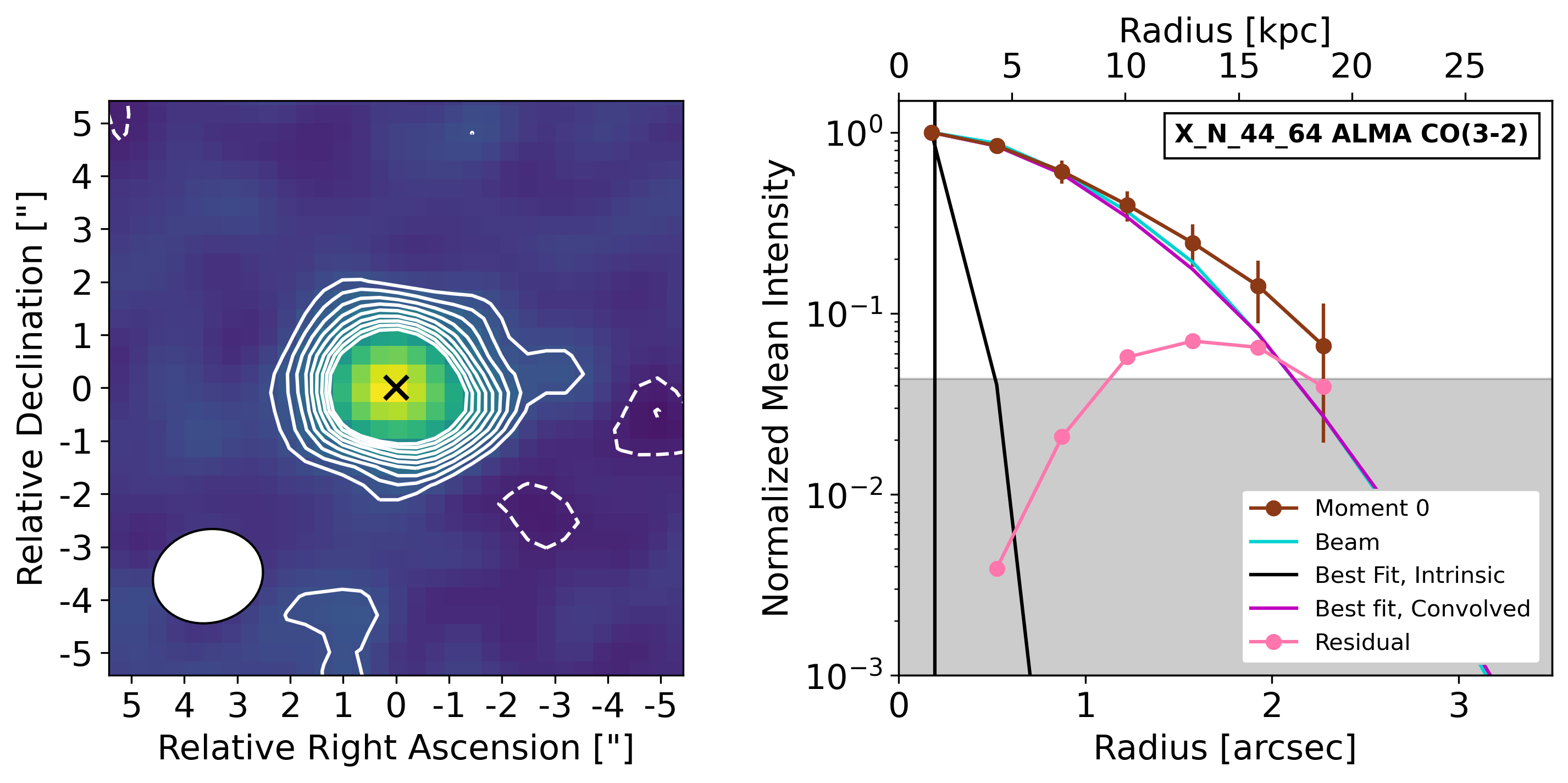}
\includegraphics[trim=11.6cm 0 0 0, clip,width=5.9cm]{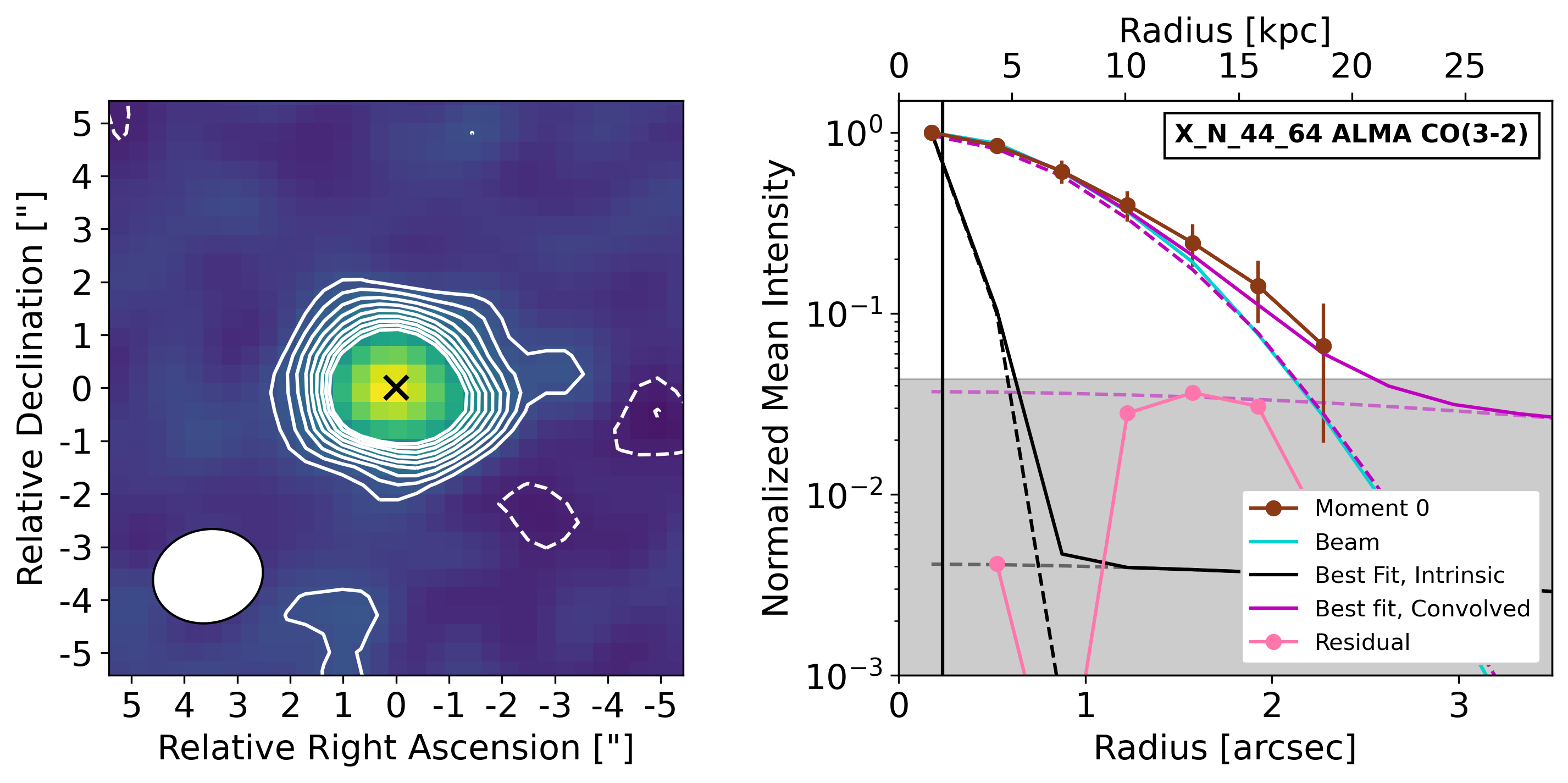}
\includegraphics[width=11.8cm]{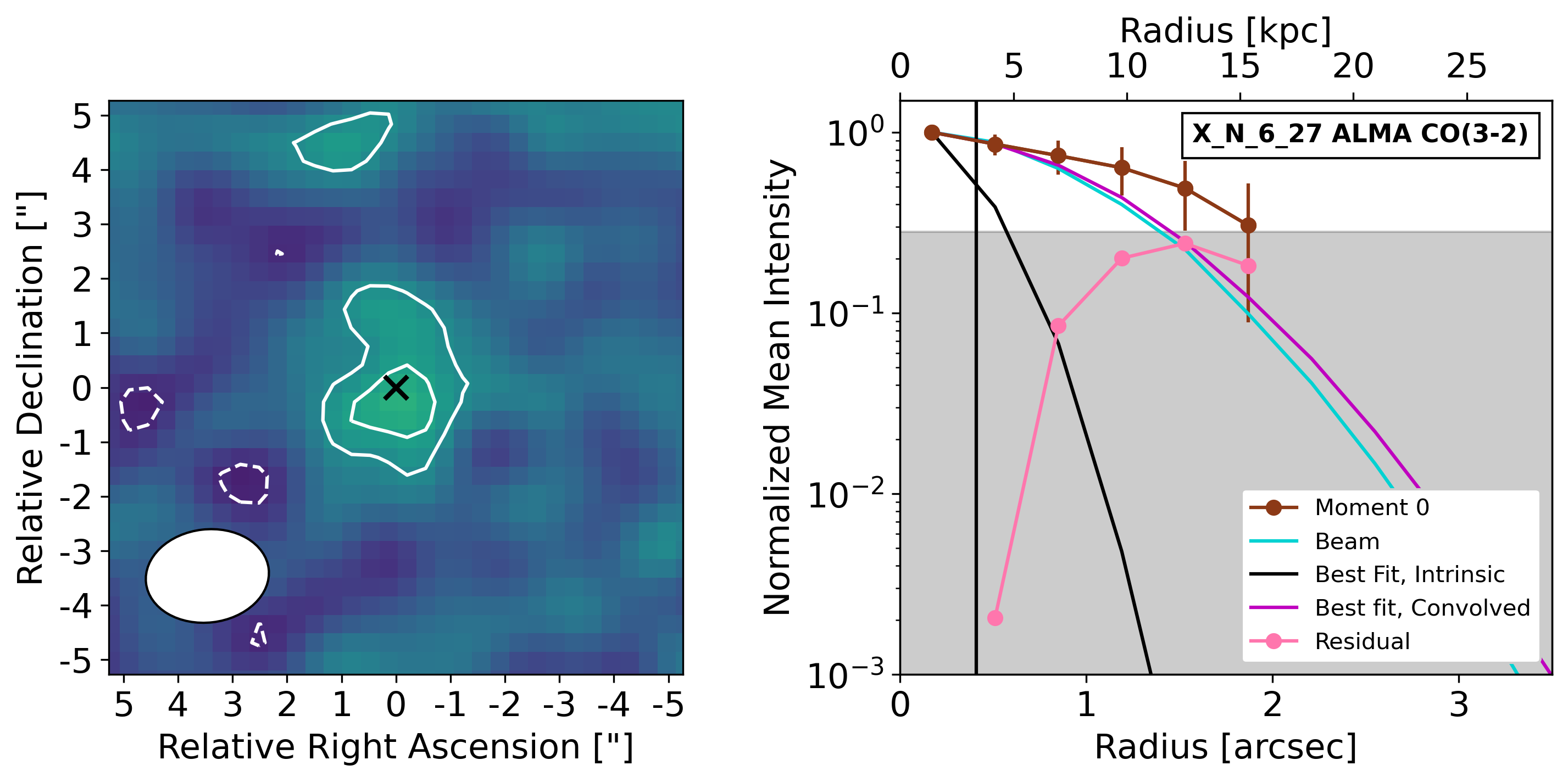}
\includegraphics[trim=11.6cm 0 0 0, clip,width=5.9cm]{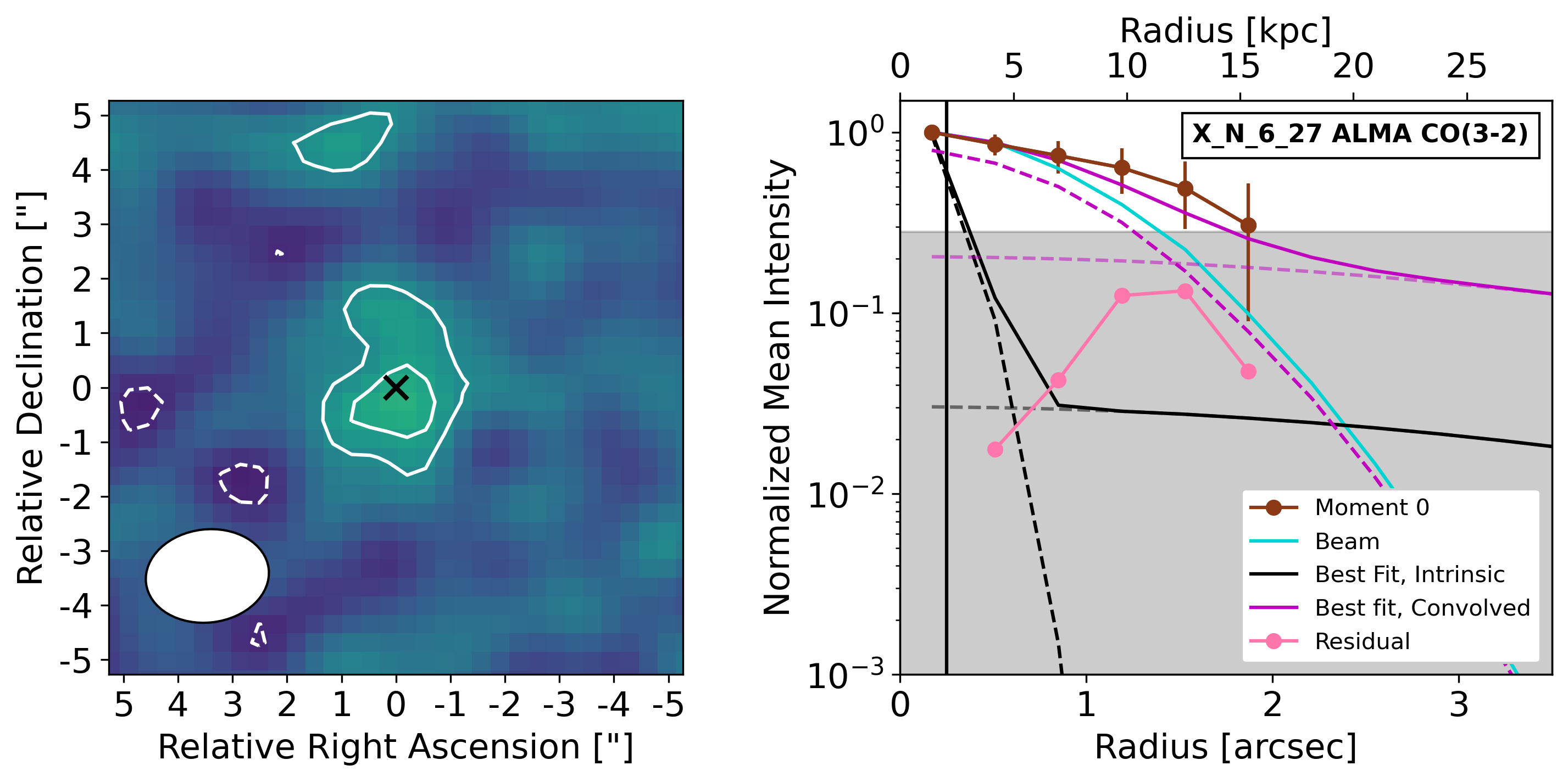}
\caption{(Cont.)}
\end{figure*}

\subsubsection{Results}
For each observation where CO(3-2) emission was significantly detected, we present the best-fit parameters and goodness of fit in Table \ref{pmntable}. The resulting profiles are shown in Figure \ref{PMN_ACA}. Below, we discuss the implications of these fits.

The goodness of the fit is characterized in three ways: $\chi^2$, $\chi_{red}^2$, and the Bayesian evidence outputted by PyMultiNest ($Z$). This last value is sometimes useful for determining which model is preferred, as a ratio of Z$_1$/Z$_2>10$ would indicate slight evidence towards model 1 (e.g., \citealt{jeff61}).

For the ACA data of \cidthree, we find that the observed radial profile of the ACA data is nearly identical to that of the synthesized beam, as seen by the overlap between the maroon and cyan curves in Figure \ref{PMN_ACA}. Allowing for a Gaussian (resolved) component results in a better $\chi^2$, but a very small source (HWHM$\sim0.5''$) that is much less than the cell size, let alone the resolution of the observation. Further allowing for an additional Gaussian component (2G) makes the fit unstable, by increasing $\chi_{red}^2$, which is what is expected when starting from a model that is already overfitting the data.  The evidence ratio $Z_{1G}/Z_{2G}$ is $\lesssim2$, further indicating that the inclusion of an additional component is not required, and that the most likely fit is an unresolved point source.

The other source detected in CO(3-2) emission in ACA data (\xnfour) shows a radial profile that extends beyond the beam, although with high uncertainties. Examining the moment zero map in Figure \ref{PMN_ACA}, it is apparent that this is partly due to the fact that there is a $\sim1\sigma$ tail of emission that extends to the west and north of this source, which shifts the best-fit centroid slightly. Despite this, the best-fit model is a point source, as seen by the goodness of fit measures.

All of these results lead to the conclusion that the ACA data in these sources is unresolved, implying that the ACA data originates from a source of radius $\lesssim6''\sim50$\,kpc. On the other hand, the ALMA data of these two sources is best fit by two Gaussian components, implying a resolved central source (HWHM$\sim 1$\,kpc) and a spatially extended component (HWHM$\sim 40$\,kpc; Figure \ref{PMN_ACA}), which is consistent with the upper limit from ACA.
%The difference in findings between the ALMA and ACA data is likely due to the difference in sensitivities (i.e., the ALMA is $\sim 3\times$ more sensitive).

While the ACA cube of \xnsix shows no significant line emission, the ALMA moment zero map is best fit by a composite source (HWHM$\sim2$\,kpc and a much weaker source with HWHM$\sim34$\,kpc). This source is composed of a small $3\sigma$ peak and a more extended $2\sigma$ that extends to the north. So it is also possible that the morphology is truly unresolved, but inflated by noise peaks.

In a parallel work \citep{jone23}, we combine the data from seven ALMA CO(3-2) observations of SUPER galaxies by performing a stacking analysis. The resulting stacked data cube, when collapsed over $\pm100$\,km\,s$^{-1}$, is best fit by a two component model with a bright central source (radius $\sim1$\,kpc) and a weaker extended component (radius $\sim14$\,kpc). These results from the stacking analysis are generally consistent with the radial profiles of the two ALMA high-sensitivity measurements of the two individual galaxies presented in this paper.
This paints a picture of two components in these galaxies: a bright central galaxy of radius $\sim1$\,kpc, and a spatially extended component of radius $\sim10-50$\,kpc, both of which are enclosed within the ACA beam.
There is no evidence of a third component at larger spatial scales (i.e., $>100$\,kpc).

\begin{table*}
 \centering
 \begin{tabular}{ccc|ccc|ccc}
Galaxy	&	Data	&	Model	&	log$_{10}($HWHM$_{\mathrm{G1}}/[$UNIT$])$	&	log$_{10}($HWHM$_{\mathrm{G2}}/[$UNIT$])$	&	log$_{10}($f$_{\mathrm{12}})$	&	$\chi^2$	&	$\chi_{red}^2$	&	ln(Z)	\\	\hline
\cidthree	&	ALMA	&	PSF							&	-							&	-				&	-	&	1.9	&	-	&	-	\\	
	&		&	1G	&	$-0.90\pm0.49$,$0.01\pm0.49$	&	-							&	-				&	2.5	&	0.5	&	$7.78\pm0.05$	\\	
	&		&	2G	&	$-0.78\pm0.32$,$0.14\pm0.32$	&	$0.69\pm0.44$,$1.61\pm0.44$ &	$-2.59\pm 0.53$	&	0.7	&	0.2	&	$7.74\pm0.06$	\\	
	&	ACA	&	PSF	&	-								&	-							&	-				&	0.3	&	-	&	-	\\	
	&		&	1G	&	$-0.34\pm0.62$,$0.58\pm0.62$	&	-							&	-				&	0.04&	0.02&	$4.97\pm0.04$	\\	
	&		&	2G	&	$-0.22\pm0.67$,$0.69\pm0.67$	&	$1.00\pm0.43$,$1.92\pm0.43$	&	$-2.11\pm0.91$	&	3.4	&	3.4	&	$4.34\pm0.05$	\\	\hline
\xnfour	&	ALMA	&	PSF								&	-							&	-				&	-	&	3.2	&	-	&	-	\\	
	&		&	1G	&	$-0.71\pm0.52$,$0.20\pm0.52$	&	-							&	-				&	4.0	&	0.7	&	$11.51\pm0.05$	\\	
	&		&	2G	&	$-0.63\pm0.30$,$0.29\pm0.30$	&	$0.69\pm0.45$,$1.61\pm0.45$	&	$-2.66\pm0.48$	&	0.8	&	0.2	&	$11.51\pm0.07$	\\	
	&	ACA	&	PSF	&	-								&	-							&	-				&	0.7	&	-	&	-	\\	
	&		&	1G	&	$-0.06\pm0.73$,$0.85\pm0.73$	&	-							&	-				&	1.3	&	0.3	&	$3.66\pm0.03$	\\	
	&		&	2G	&	$-0.59\pm0.76$,$0.33\pm0.76$	&	$0.93\pm0.37$,$1.84\pm0.37$	&	$-1.79\pm0.95$	&	1.6	&	0.8	&	$3.91\pm0.04$	\\	\hline
\xnsix	&	ALMA	&	PSF	&	-						&	-							&	-				&	4.7	&	-	&	-	\\	
	&		&	1G	&	$-0.39\pm0.59$,$0.53\pm0.59$	&	-							&	-				&	3.5	&	0.7	&	$4.11\pm0.05$	\\	
	&		&	2G	&	$-0.60\pm0.51$,$0.31\pm0.51$	&	$0.61\pm0.46$,$1.53\pm0.46$	&	$-1.80\pm0.91$	&	1.1	&	0.4	&	$4.95\pm0.05$			
 \end{tabular}
 \caption{Best-fit values for a one- and two-Gaussian 2D model applied to the moment 0 map of each CO(3-2) detection (see Section \ref{HALOMOD}). For each HWHM entry, we first list the value with units of arcseconds, then with units of kiloparsecs. We also note the goodness of fit through three criteria: $\chi^2$, $\chi_{red}^2$, and the natural logarithm of the Bayesian evidence, as output by MultiNest \citep{fero09}.}
 \label{pmntable}
\end{table*}

\section{Discussion}\label{DISC}

\subsection{Is there extended CO flux?}\label{extendy}
In the previous Section, we detailed the detection of CO(3-2) emission from two $z\sim2.2$ AGN host galaxies using the ACA, and also discussed the upper limit for a third source. The CO(3-2) emission of these sources had previously been detected using the 12\,m-array of ALMA \citep{circ21} and, in the case of \xnfour and \xnsix also re-observed with ALMA (this work). Since the new observations feature a synthesized beam that is $\sim10\times$ larger, it is possible that they have captured emission from spatially extended emission that would not be detected using the 12\,m array. 

\begin{figure}
\centering
\includegraphics[width=0.5\textwidth]{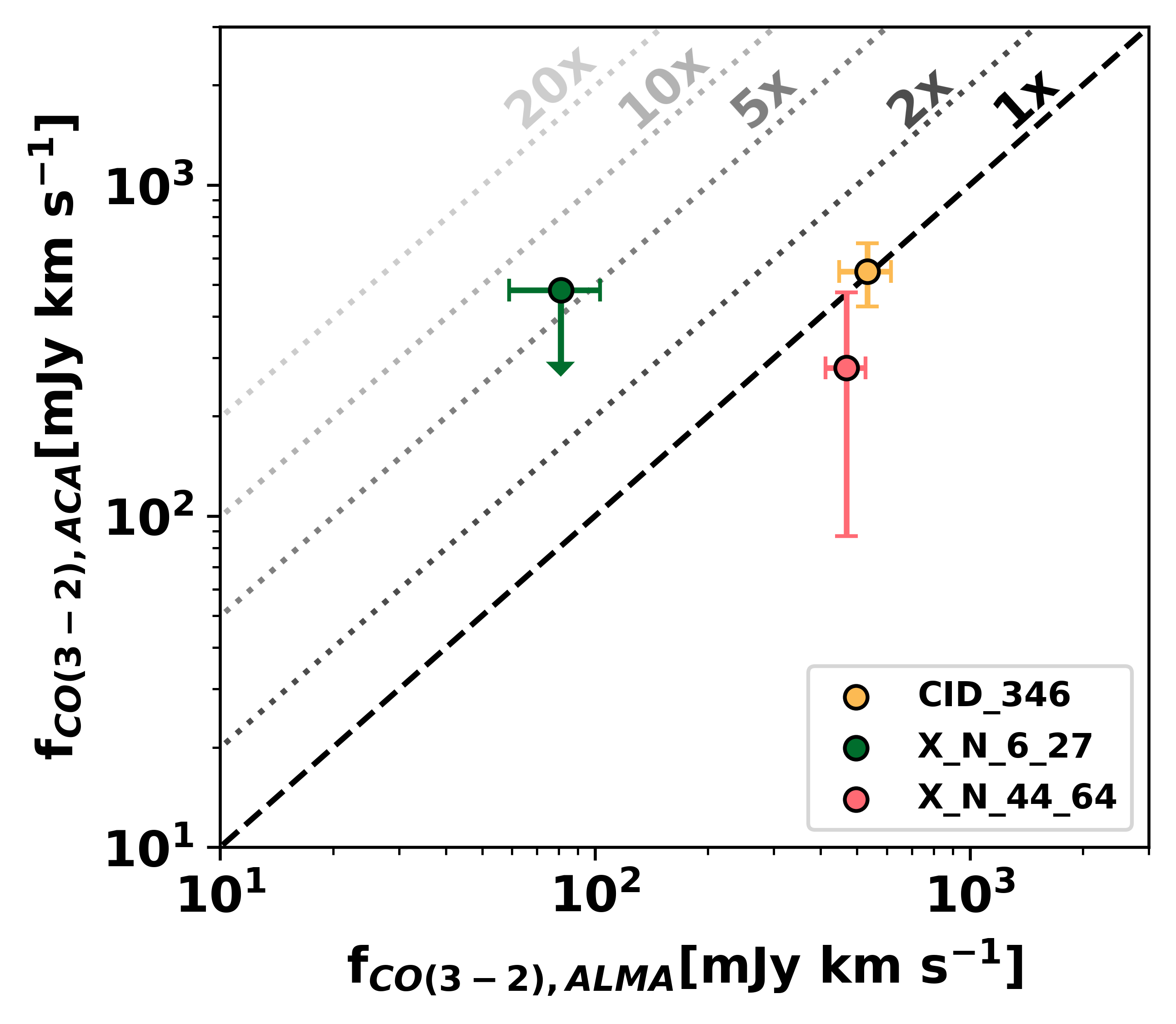}
\caption{Comparison of recovered integrated CO(3-2) flux density for each source, as measured in low-resolution ACA observations (vertical axis) and relatively high-resolution ALMA observations (horizontal axis). Since \xnsix in undetected in line emission with the ACA, the $3\sigma$ upper limit on the integrated flux density is shown by a downwards-facing arrow. Dotted and dashed lines show constant ratios of ACA flux to ALMA flux. For reference, the ALMA data have an angular scale of $\sim10$\,kpc, while the ACA data traces larger-scale emission ($\sim100$\,kpc; see Table \ref{obstable}).}
\label{ALMAvsACA}
\end{figure}

In Figure \ref{ALMAvsACA}, we compare the integrated CO(3-2) fluxes of our three sources using the ACA (vertical axis) and ALMA (horizontal axis). The two detections (\cidthree and \xnfour) show ACA fluxes that are in agreement with the observed ALMA flux. The third source (\xnsix) is undetected with the ACA, but its $3\sigma$ upper limit is $\sim5\times$ the measured ALMA flux density.

%We note that the high upper limit of \xnsix is primarily due to the poor sensitivity of our ACA observations, which result in a high $2\sigma$ upper limit. Similarly, the low CO(3-2) flux density of \xnfour is likely not physical, but is a result of our observations featuring high noise levels that hide the underlying complex spectral behaviour (i.e., a two-horned profile, as seen in the ALMA data).

At face value, this implies that the two ACA-detected sources have no evidence for extended CO emission beyond the largest recoverable scale of ALMA, while the ACA-undetected source may potentially have a halo. Indeed, our image- and visibility plane analyses show no evidence for larger recovered CO fluxes for the ACA observations. Additionally, the finding that the ACA data show no evidence of extended emission beyond the synthesized beam excludes the presence of a significant amount of molecular gas on scales larger than $\sim 60$\,kpc in radius. Together, our observations exclude the presence of massive molecular halos on scales larger than about 20\,kpc in radius.

However, it is important to note that while the ACA observations detailed here are sensitive to larger spatial scales, they also feature lower sensitivities than the ALMA observations. For example, the ACA and ALMA data cubes for \xnsix had beam sizes of $\sim10''$ and $\sim1''$, but RMS noise levels per channel of $1.4$\,mJy\,beam$^{-1}$ and $0.19$\,mJy\,beam$^{-1}$, respectively. It is therefore possible that some weak level of spatially extended emission is present between the ALMA larger recoverable scales and the ACA beam (at the level of 10-20\% of the ACA flux, see Figure \ref{ALMAvsACA}). As well, it is possible that weak spatially extended emission beyond the ACA beam is present (at the level of a few \% of the ACA peak).

\subsection{Comparison with previous results}\label{compto}

In this section we compare our sensitive ACA upper limits on the presence of molecular gas in galactic halos, on scales larger than about 20\,kpc, with previous results.

The first, most direct comparison should be made with \cite{cico21}, as they performed a similar analysis specifically on \cidthree by using ACA CO(3-2) data. This previous work presented the detection of a molecular halo with a radius of 200\,kpc. By combining this previous data with new observations, we find evidence for the CO emission in the ACA data to be unresolved (i.e. not larger than $\sim50$\,kpc in radius; see Figure \ref{uvcont_fig} and Figure \ref{PMN_ACA}). Furthermore, while this previous work found the CO flux in the ACA data to be 14 times higher than what is observed with ALMA, our combined dataset shows that the ACA CO flux is fully consistent with the ALMA CO flux (Figure \ref{ALMAvsACA}).

The origin of such a discrepancy is not clear. Our combined ACA data are deeper than the dataset analysed in \cite{cico21}. Since the combined data represent a factor of $\sim2.3$ longer on-source integration time\footnote{We note that the integration time reported in Table 2 of \citet{cico21} is the total execution time (5.2\,hours), while the on-source integration time is 3.3\,hours.}, the resulting sensitivity should be increased by a factor of $\sim1.5$\footnote{Using the ALMA sensitivity calculator; \url{https://almascience.eso.org/proposing/sensitivity-calculator}.}. Due to different beam sizes and number of antennas, our sensitivity per channel is increased by a factor of $\sim1.4$. However, this difference in sensitivity is not enough to explain this difference in findings.

\begin{figure*}
    \centering
    \includegraphics[width=0.6\textwidth]{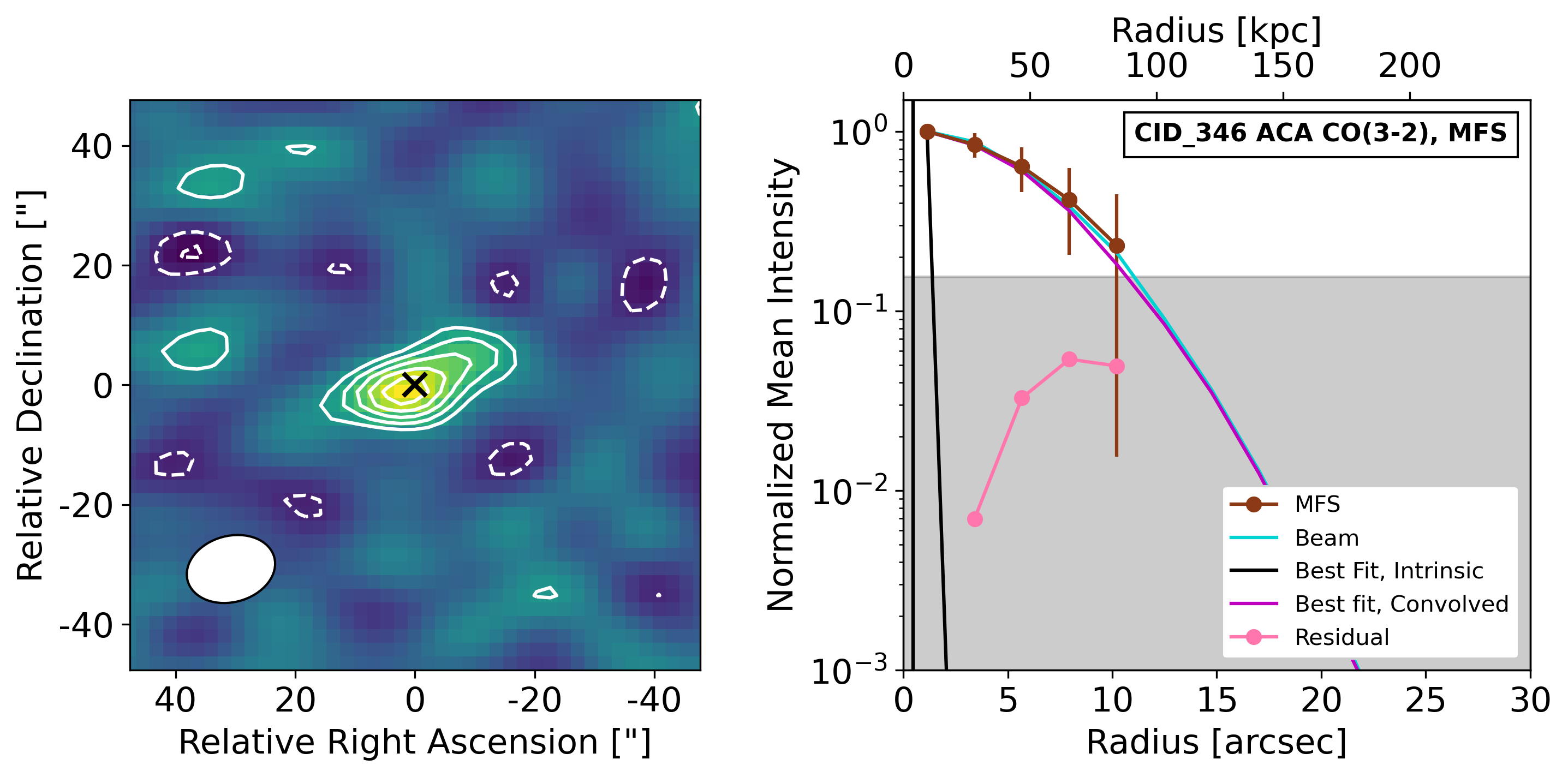}
    \caption{Results of fitting a single-component, two-dimensional Gaussian model to the MFS image of CO(3-2) emission from \cidthree with no PB correction, multiplied by the velocity bandwidth used. The left panel shows a zoomed in intensity map, to be compared with the moment zero map of Figure \ref{mom0_6}. The contours are shown at $\pm(2,3,4,\ldots)\sigma_{RMS}$ (where $1\sigma_{RMS}=0.09$Jy\,beam$^{-1}$\,km\,s$^{-1}$). Synthesized beam and galaxy position given by ellipse and cross, respectively. }
    \label{cid346spec2}
\end{figure*}

\cite{cico21}  extracted the integrated CO map using the MFS mode to collapse the channels covered by the line (V. Mainieri, priv. comm.). We have employed the same method on our data (see Figure \ref{cid346spec2}), which does not show any significant difference with respect to the moment 0 map extracted in this work -- the noise is slightly lower, but the size and flux are fully consistent with those derived in the previous sections. Due to the lower noise level, a weak spatial extension to the northwest appears to emerge, but with the same elongation as the beam. As seen in the right panel of Figure \ref{cid346spec2}, the radial profile of this emission is still very well fit by the PSF. Since the extension is unilateral and has a low significance, we interpret it as not significant.

An additional possible origin of this disagreement is that this previous work uses a different velocity integration interval ($\rm -400<v<1000$\,km\,s$^{-1}$) based on the broad and asymmetric CO profile that was obtained. We do not obtain such a broad and asymmetric profile (Fig.\ref{spec_6}), regardless of the extraction aperture that we adopt (see Appendix \ref{acaspec} for a further exploration of this).
%If we obtain a map by adopting the same broad and asymmetric integration velocity interval, \textbf{we notice that the S/N of the detection decreases and that the peak is slightly off-centered, in the direction of the beam elongation (see Appendix A). However, even in this case we do not obtain the huge.}
However, in Appendix \ref{cicone} we repeat the same analysis by using the large, asymmetric velocity range adopted by \cite{cico21}: the peak is slightly off-centered, but we do not obtain any evidence of the large extension that was previously obtained.

Possible differences may arise during steps of image creation (``cleaning'') that were not specified by \citet{cico21}. First, it appears that this previous work used very small image cells ($\sim 1''$, or 1/10$^{\mathrm{th}}$ the FWHM of the minor axis of the restoring beam) that may over-resolve the emission (see discussion in Appendix \ref{appb}). In addition, a difference may arise when adopting a low threshold when performing the cleaning (e.g., $0\sigma$ rather than $3\sigma$), which could result in an artificially boosted flux within the cleaning aperture, especially if using interactive cleaning. Something similar could have happened to the data published by \citet{cico21} and explain the different results. 

Within this context it is interesting to note that the \textit{uv} radial profile shown by \citet{cico21} is fully consistent with ours within the uncertainties. This previous work reported a very low significance detection, which agrees with our non-detection. 

With the exception of \cite{cico21}, our results are broadly consistent with the results obtained by other authors on cold halos around high-redshift galaxies and quasar host galaxies. Indeed, as mentioned in Section \ref{intro}, studies based on tracing the CO and [CII] transitions in galaxies and quasar at $z\sim2-6$ have resulted in the detection of halos on scales of order $\sim10$\,kpc, and with a mass that is lower or comparable to the mass of cold gas in the ISM of the central galaxy
\citep{Ginolfi17,fuji19,fuji20,Ginolfi20,herr21}.

In a parallel work we stack the CO data of seven SUPER-ALMA galaxies and obtain consistent results with this work (i.e., weak CO halos on scales of $\sim 10-20$\,kpc; \citealt{jone23}). In an additional work \citep{scho23}, we investigate the halos of powerful extremely red quasars at $z\sim2.5$ by using the [CI] transition (a more reliable tracer of molecular gas in some environments; e.g., \citealt{dunn21,jiao21}) and find the same result (i.e. the existence of weak cold halos, but on scales of $\sim 10-20$\,kpc).

The properties of the cold gas in the CGM are likely different in the case of radio galaxies, for which molecular gas emission is indeed observed on scales of $\sim100$\,kpc \citep{Emonts16,Li21}. In these cases, the CGM enrichment with molecular gas may result from ISM lifting from the radio jets or compression and cooling of the CGM in the expanding radio lobes or cocoons 
\citep{Russell17,Russell19}.

\section{Conclusions}\label{conc}
In this work, we have presented new ACA and ALMA observations (and combined archival data) of CO(3-2) emission around three AGN host galaxies at cosmic noon to search for evidence of an extremely extended cold gas in the CGM (i.e., of order $\sim100$\,kpc). The combination of ACA and ALMA data allows the comparison of small- and large-scale emission.
\begin{itemize}

\item Using the ACA data alone, no rest-frame FIR continuum emission ($\lambda_{rest}\sim870$\,$\mu$m) is detected from each individual object in the image plane, and a stack of the continuum visibilities from all three galaxies shows no significant signal. Due to the higher sensitivity of the ALMA observations, continuum emission from two of these three sources is detected in ALMA data at the same frequency. The upper limits on large-scale continuum flux densities for each source are in agreement with the ALMA values, so we find no evidence for extended continuum emission in these sources.

\item By stacking the continuum visibilities, we find the tentative detection ($\sim2\sigma$) of negative continuum emission on large scales ($>500$\,kpc). This feature could be tracing the thermal S-Z effect associated with the halo heating from AGN feedback. The level of the tentative signal is consistent with the expectations of cosmological simulations for this important phenomenon and will be followed up with future observations.

\item CO(3-2) emission is detected with ACA in two sources (\cidthree and \xnfour). All three sources are detected in CO(3-2) emission with ALMA, and the ALMA and ACA flux densities are in agreement. This indicates that there is no evidence for CO emission extended beyond ALMA's largest recoverable scale that is missed by ALMA, within the ACA beam.

\item To search for CO emission extending beyond the ACA beam, we investigate the CO emission in the ACA data both through the analysis of radial brightness profiles in the maps and by analysing the visibilities.  We find that the CO(3-2) emission in the ACA data of \cidthree  and \xnfour is consistent with being unresolved ($r<40-50$\,kpc). On smaller scales, the ALMA data for these two sources reveals the presence of two components: a compact ($r<1-2$~kpc) component and a weak, extended component ($r\sim10-50$\,kpc).
%The ALMA data for \xnsix hints at an extended source ($r\sim3$\,kpc), but this may be an effect of noise.

%\item To compare with previous works, we also examine the effects of using a different spectral integration scheme (i.e., MFS rather than moment 0). While MFS images of line emission result in slightly lower noise levels, the resulting emission morphology is not significantly affected for our galaxies.  

\end{itemize}

%We re-analyse the previous \cidthree ACA CO(3-2) data, which was claimed to feature an extension to $r\sim$\,kpc. By following a similar same calibration and imaging procedure as the previous work, we find a point source with no CO halo. This different conclusion is a result of assuming a more conservative cleaning procedure (i.e., automatic cleaning with a set $3\sigma$ threshold) and the examination of fitting uncertainties in our visibility analysis.

\noindent
Altogether, our results show that these $z\sim2.2$ AGN hosts galaxies feature CO(3-2) spatial extents of $r\sim10-50$\,kpc (with a more concentrated component on scales of $r\sim 1$~kpc, likely tracing the ISM of the star forming galaxy), with no evidence of $r\sim100$\,kpc components. In our companion paper \citep{jone23}, we similarly find that the high-resolution (i.e., ALMA) CO(3-2) morphologies imply a compact central source (co-spatial with the central, star forming galaxy; $r\sim1$\,kpc) and an extended component (the CGM; $r\sim10$\,kpc). Thus, it is likely that these sources are surrounded by a gaseous reservoir $\sim10\times$ larger than the galaxy, with no evidence for an additional component $\sim100\times$ larger than the galaxy. 

Our results are consistent with other studies of galaxies and quasars at $z\sim2-6$, which find cold halos on similar scales, but are inconsistent with recent discoveries of huge amount of cold gas in the CGM on scales of 200\,kpc or larger. On the other hand, the evidence for molecular gas on $\sim100$\,kpc scale appears solid around radio galaxies/radio-loud quasars \citep{Emonts16,Li21}, where ISM jet lifting of CGM/ICM compression may be responsible for the presence of molecular gas on very large scales.

Finally, the tentative detection of the S-Z signal is extremely intriguing. If confirmed at high significance with deeper observation this would be one of the very first direct evidences of halo heating from the AGN feedback, which is predicted by models to be an important phenomenon preventing cold accretion onto the galaxy and resulting into quenching as a consequence of starvation. However, deeper ACA observations and on a larger sample of AGN at high-z are certainly needed both to confirm the detection and to explore whether this is an ubiquitous phenomenon.

\section*{Acknowledgements}
We are grateful to Chiara Feruglio and Celine Peroux for useful comments.
This paper is based on data obtained with the ALMA Observatory, under programs 2016.1.00798.S, 2019.2.00118.S, and 2021.1.00327.S. ALMA is a partnership of ESO (representing its member states), NSF (USA) and NINS (Japan), together with NRC (Canada), MOST and ASIAA (Taiwan), and KASI (Republic of Korea), in cooperation with the Republic of Chile. The Joint ALMA Observatory is operated by ESO, AUI/NRAO and NAOJ. 
G.C.J. acknowledges funding from ERC Advanced Grant 789056 ``FirstGalaxies’’ under the European Union’s Horizon 2020 research and innovation programme. R.M. and J. S.  acknowledge funding from ERC Advanced Grant 695671 ``QUENCH’’, as well as support by the Science and Technology Facilities Council (STFC). R. M. additionally acknowledges the support from a Royal Society Research Professorship. S.C is supported by European Union’s HE ERC Starting Grant No. 101040227 - WINGS. Y.F. acknowledge support from NAOJ ALMA Scientific Research Grant number 2020-16B.

\section*{Data Availability}
The data analysed in this work are available from the ALMA data archive (\url{https://almascience.nrao.edu/asax/}).

\bibliographystyle{mnras}
\bibliography{references}

\appendix

\section{Statistical visibility analysis}\label{appb}

In Section \ref{ipco} of the main text, we explore the CO(3-2) visibilities of the combined ACA observations of \cidthree, assuming a small velocity range (i.e., $\pm$HWHM, or $ -91<$v$<91$\,km\,s$^{-1}$). By fitting multiple models to the visibilities, we find a strong detection of a point source, with no evidence for extended emission.

At first, this finding may appear to contradict the conclusion of a previous work (\citealt{cico21}), which found an extended component in the CO visibilities of a single ACA observation. However, the velocity range of the previous work was much more extended ($-400<v<1000$\,km\,s$^{-1}$), and we will show that this emission does not show significant extension in Section \ref{appa}. Here we apply a statistical fitting analysis to the same CO visibilities to show that there is no significant ($>3\sigma$) extended emission.

\begin{figure}
\centering
\includegraphics[width=0.5\textwidth]{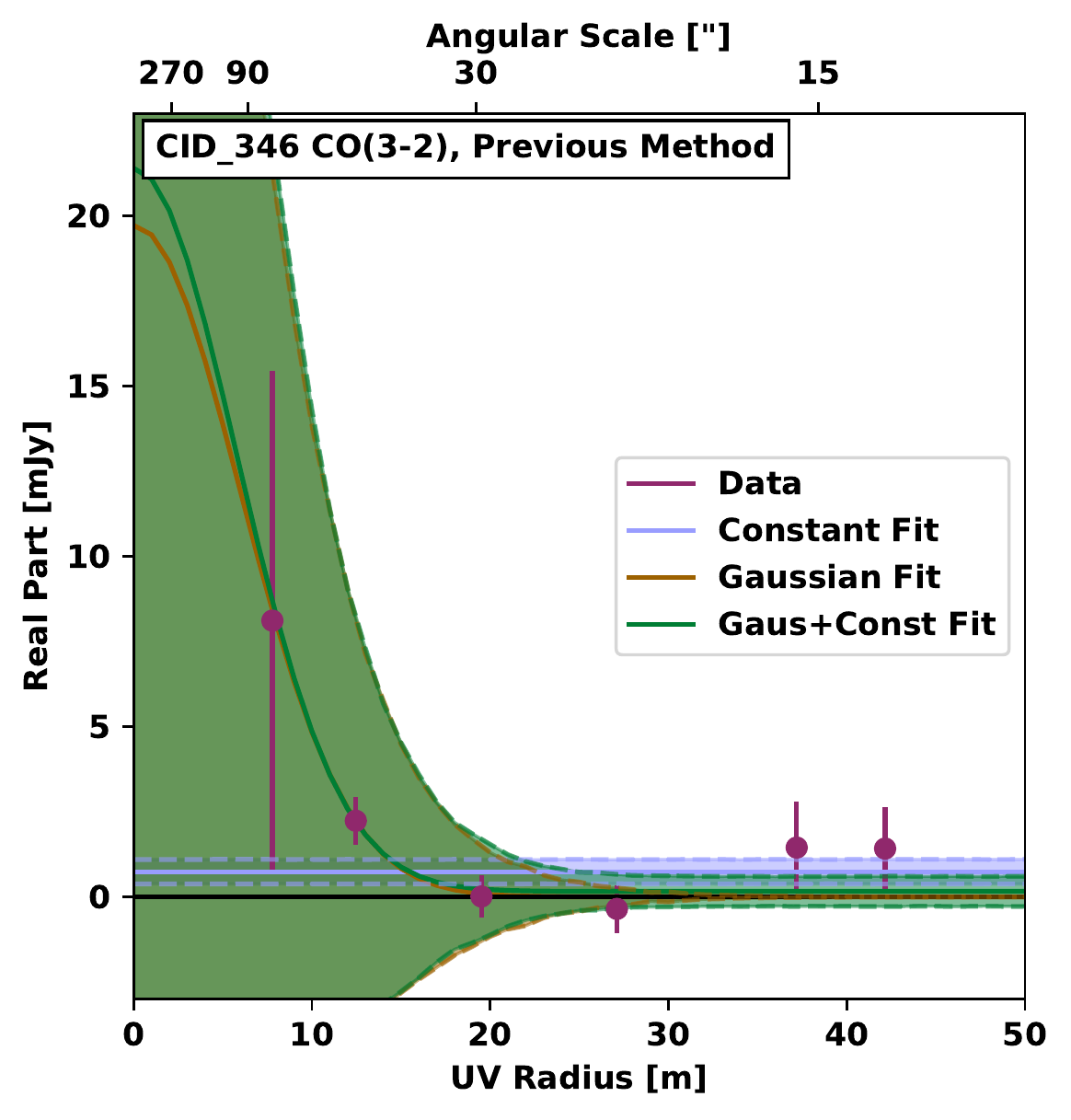}
\includegraphics[width=0.5\textwidth]{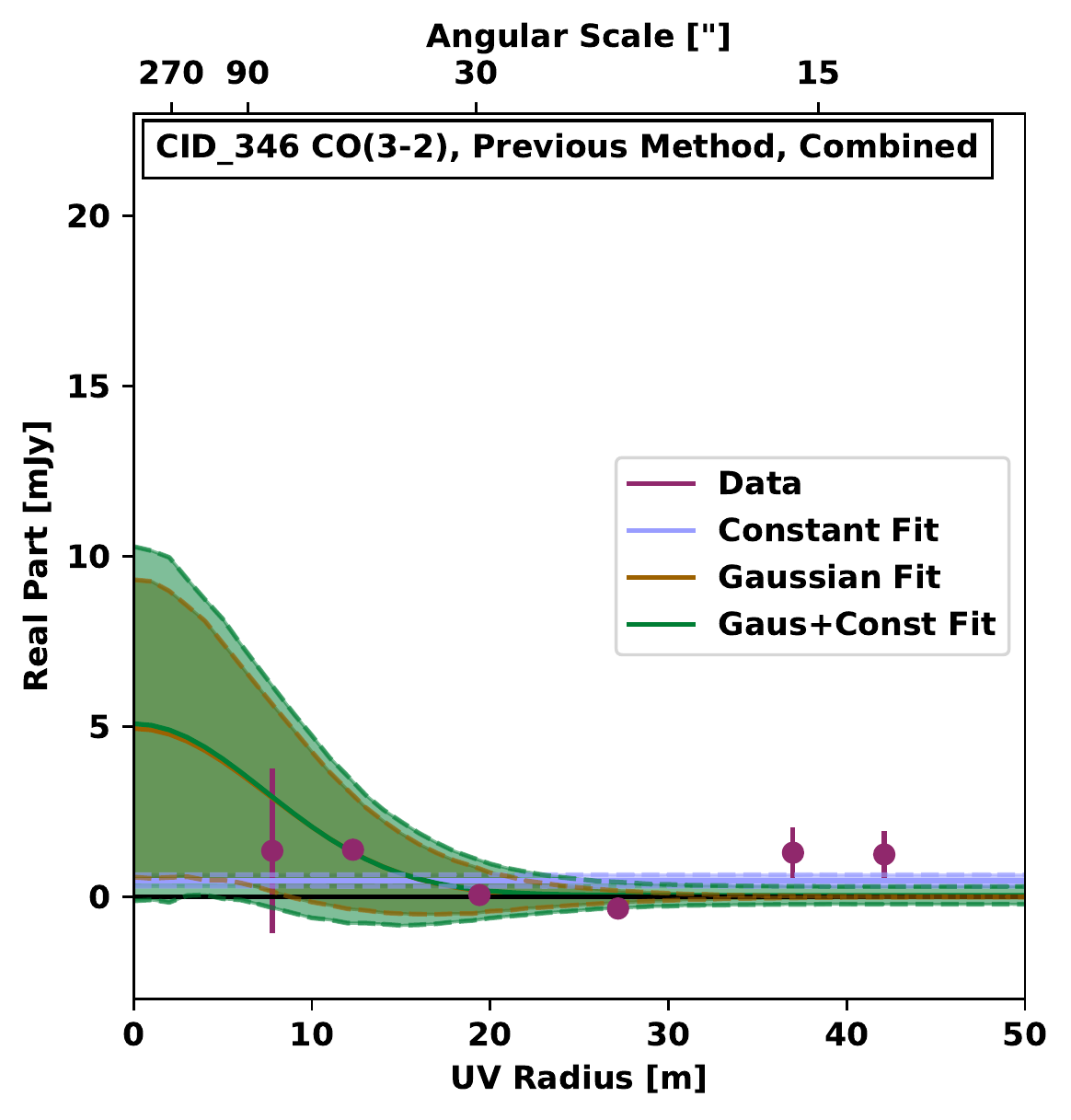}
\caption{Radial profile of the real part of the observed visibilities of the \citet{cico21} data (top panel) and the combined ACA data for \cidthree (lower panel), including only channels in the range $-400<v<1000$\,km\,s$^{-1}$ (maroon points). The results of fitting three models are shown: a single Gaussian (brown line), a constant value (light blue line), and an offset Gaussian (green line). Uncertainties shown by shaded regions ($1\sigma$). Note that the uncertainty regions for the two Gaussian models (orange and green) are nearly identical.}
\label{threefit}
\end{figure}

We first extract the CO(3-2) visibilities for \cidthree from project 2019.2.00118.S (i.e., all channels between $-400<v<1000$\,km\,s$^{-1}$), place them into \textit{uv}-bins, and plot them using the \textit{uvplot} package \citep{tazz17}. By using the same \textit{uv} bins as \citet{cico21} (8\,m), we find a similar profile to that of \citet{cico21}. We then fit three models to this profile: a constant value, a Gaussian, and an offset Gaussian. In image space, these correspond to a point source, a resolved source, and a coincident resolved source and point source, respectively (see Figure \ref{threefit}). We then repeat this process with the combined dataset of ACA data for \cidthree.

To check the goodness of fit, we calculate the $\chi^2$ and $\chi_{red}^2$ value of each fit. As shown in Table \ref{vistable_C}, models with Gaussian components (hence with a resolved component of the CO emission) better fit the data. However, we note that the two Gaussian fits to the `Original' dataset give amplitudes that are within $1\sigma$ of 0; so there is no detection of any resolved CO component, hence no detection of extended CO emission. This is shown by the large uncertainty regions of the the top panel of Figure \ref{threefit}.
Since the best model returns a constant amplitude of $0.74\pm0.36$ (i.e., $\sim2\sigma$),
we conclude that this data (when collapsed over the given, large velocity range) contains no significant evidence for either a point source or an extended component, in agreement with Section \ref{appa}.

The `Combined' dataset contains more data, so the errors in the visibility profile are smaller (see lower panel of Figure \ref{threefit}). Since the lowest-\textit{uv} point is now much lower in amplitude, the evidence for an extended component is decreased. As for the `Original' dataset, the Gaussian fits return better goodness-of-fit values and unconstrained amplitudes, while the constant fit returns a $<3\sigma$ value. Therefore, these \textit{uv} data show little evidence for significant emission. We note that an MFS image of the `Combined' visibilities results in a $\sim3\sigma$ central peak (Figure \ref{fourmfs_1}), so there may be a weak signal in the visibilities that is occluded by the inclusion of noisy data (as seen by the multiple $\pm2\sigma$ noise peaks in the image).

\begin{table*}
    \centering
    \begin{tabular}{cc|cccc|cc}
& & Constant & Gaussian & Gaussian & Gaussian & &\\
Dataset & Model & Amplitude & Amplitude & $c$ & $c$    & $\chi^2$ & $\chi_{red}^2$\\   
  &    & [mJy]      & [mJy]    & [m] & [$''$] &  \\ \hline \hline
Original & Constant Fit & $0.74\pm0.36$ & $\times$ & $\times$ & $\times$ & 9.89 & 1.98\\
&Gaussian Fit & $\times$ & $19.71\pm27.03$ & $5.96\pm1.96$ & $97^{+47}_{-24}$ & 2.77 & 0.69\\
&Constant+Gaussian Fit & $0.16\pm0.44$ & $21.23\pm29.97$ & $5.76\pm1.91$ & $100^{+50}_{-25}$ & 2.63 & 0.88\\ \hline
Combined&Constant Fit & $0.48\pm0.17$ & $\times$ & $\times$ & $\times$ & 19.65 & 3.93\\
&Gaussian Fit & $\times$ & $4.95\pm4.38$ & $7.54\pm2.47$ & $76^{+37}_{-19}$ & 8.33 & 2.08\\
&Constant+Gaussian Fit & $0.05\pm0.26$ & $5.04\pm5.24$ & $7.38\pm3.06$ & $78^{+55}_{-23}$ & 8.29 & 2.76\\
    \end{tabular}
    \caption{Best-fit values for three models applied to the visibility profile in Figure \ref{threefit}. We also note the goodness of fit using two statistics.}
    \label{vistable_C}
\end{table*}

\section{Updated ACA Spectrum for \cidthree}\label{acaspec}

Previous analysis of the CO(3-2) emission in \cidthree showed evidence for emission over a broad velocity range ($-400<v<+1000$\,km\,$^{-1}$) when a spectrum is extracted over a large circular aperture ($r\sim15''\sim124$\,kpc; see figure A.3 of \citealt{cico21}). By adding the new ACA data for this object, we are able to create a deeper data cube and extract a comparable spectrum. 

\begin{figure}
    \centering
\includegraphics[width=0.5\textwidth]{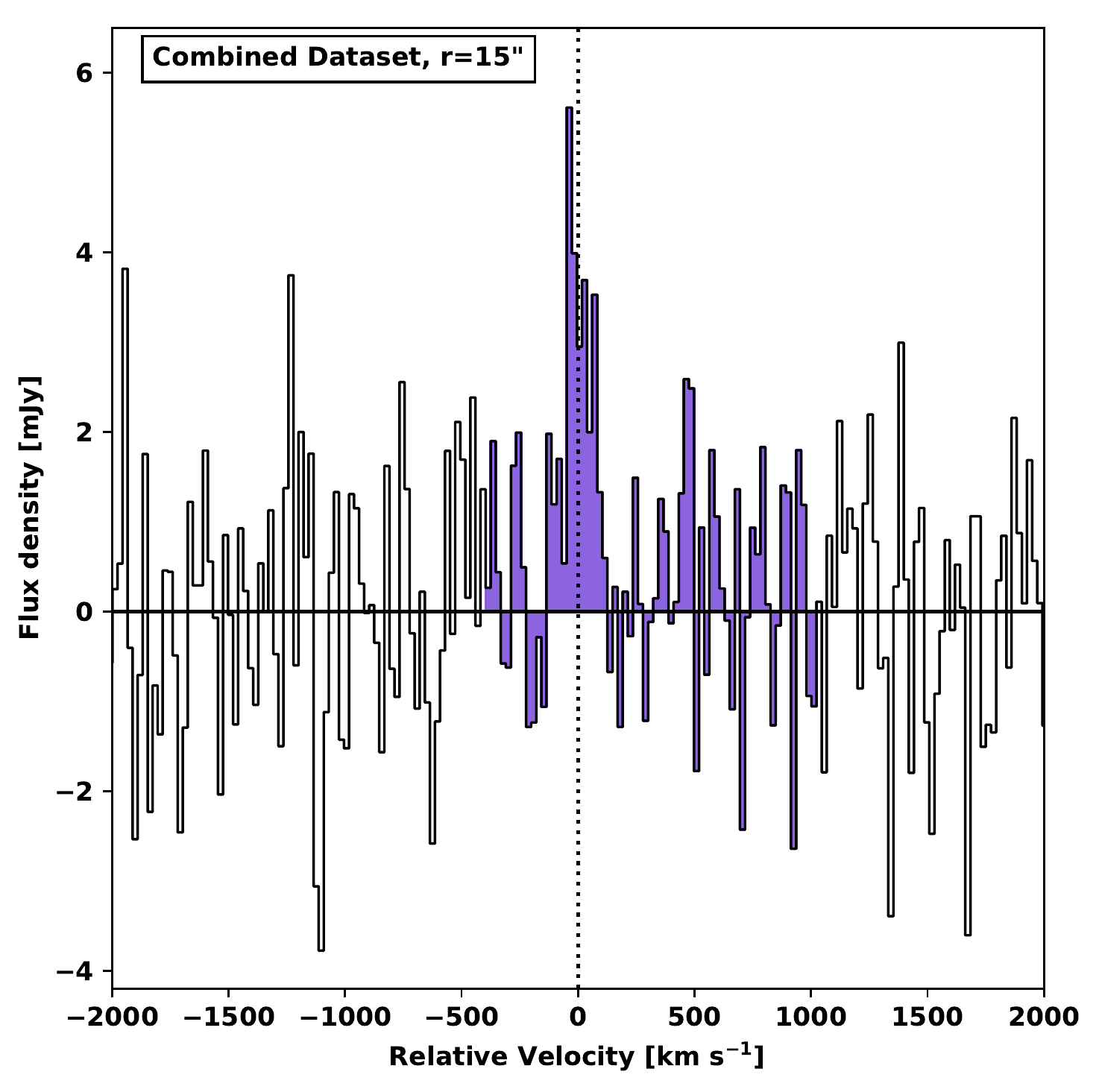}
    \caption{Spectrum of \cidthree extracted from the combined ACA dataset using a circular aperture of radius $15''$. Channels within the velocity range $-400<v<+1000$\,km\,s$^{-1}$ are highlighted in purple.}
    \label{MAR9}
\end{figure}

As seen in Figure \ref{MAR9}, the emission is concentrated at low velocities ($|v|\lesssim100$\,km\,s$^{-1}$). The amplitude of the central peak is slightly less than is seen in Figure \ref{spec_6}, and is comparable to that of the previous analysis. While there is a slight trend for the high-velocity channels, the possible signal lies within the noise level.

\section{Effects of imaging parameters on MFS maps}\label{cicone}
%A relatively high-resolution ALMA study of the CO(3-2) emission of the $z=2.2198$ BL AGN host galaxy \cidthree revealed a strong, compact detection \citep{circ21}. Recently, \citet{cico21} analysed lower-resolution ACA CO(3-2) data for \cidthree. Since the ACA is more sensitive to large-scale emission than ALMA, these observations have the potential to reveal the presence of a halo out to very large radii. Indeed, \citet{cico21} report evidence for a halo out to $r\sim 200$\,kpc, over a broad velocity range ($-400<v<+1000$\,km\,s$^{-1}$).
Here, we explore how the finding of extended CO emission obtained by \cite{cico21} may have been influenced by the choice of imaging parameters selected in the CASA task \textit{tclean}. We also investigate whether the deeper ACA data for \cidthree presented in this paper show signs of spatial extension when imaged over the same velocity range adopted by \cite{cico21}.

The image creation procedure of \citet{cico21} is followed as closely as possible. Data were downloaded from the ALMA archive and calibration was applied by running the \textit{scriptforpi.py} prepared by observatory staff. No continuum subtraction, channel averaging, or primary beam correction was applied. Uncalibrated edge channels are excluded from analysis. Line emission is assumed to lie in the velocity range $-400<v<+1000$\,km\,s$^{-1}$ of the systemic redshift ($z=2.2197$).

Using the calibrated measurement set, we first apply the CASA task \textlcsc{tclean} in MFS mode to create a dirty image using a primary beam cutoff of $20\%$ and Briggs weighting with a robust value of 0.5. While the previous work performed interactive cleaning, we do not find strong emission, and thus proceed with the dirty images (but see note on interactive cleaning in Section \ref{compto}).

As a next step, we must determine the image size and cell size. The image size is important when considering ACA observations, where the ratio of the synthesized beam to the primary beam is high (i.e., with respect to ALMA observations). In cases like this, small images may be dominated by the central source, resulting in biased noise levels. Larger images, which extend to the primary beam limit, allow for better noise estimates.

On the other hand, cell sizes affect all interferometric observations. A general rule of thumb is to set the cell size so that there are $\sim 3-5$ pixels across the smallest width of the beam (e.g., \citealt{stan19,dacu21,mich21}), although some works use $\sim 5-10$ across the beam (e.g., \citealt{nguy20,pens21,garc22}). While a smaller cell size results in a map with a higher apparent resolution, the emission (and noise) is truly correlated on scales of the beamsize.

To explore these effects we created MFS maps using two image sizes (100\,px and 256\,px) and two cell sizes ($1''$ or $\sim10\%$ of the beam, and $2''$ or $\sim20\%$ of the beam). Note that by default, CASA assumes image and cell sizes of 100\,px and $1''$, respectively.

\subsection{Effects on single measurement set}\label{appa}
\begin{figure*}
    \centering
    \includegraphics[width=0.6\textwidth]{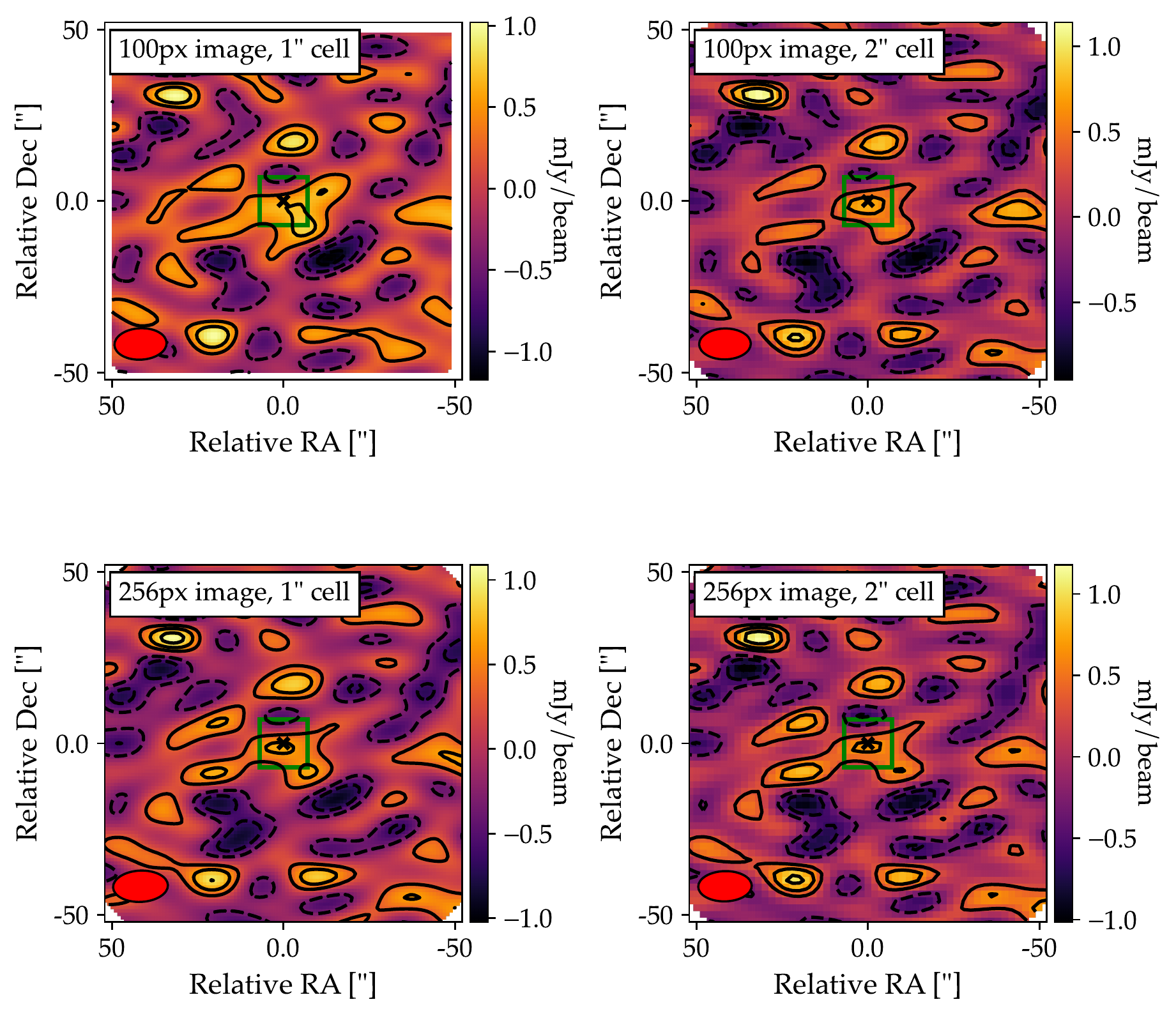}
    \caption{MFS images of CO(3-2) emission in \cidthree from ACA data. In each panel, we use the same dataset (2019.2.00118.S) and velocity range ($-400<v<1000$\,km\,s$^{-1}$), with Briggs weighting (robust parameter of 0.5). However, we vary the cell size ($1.0''$ in the left column, $2.0''$ in the right column) and field of view ($100\times100$\,px in the top row, $256\times256$\,px in the bottom row). The contours are shown at $\pm[1,2,3,\cdots]\sigma$, where $1\sigma=[0.35,0.31,0.31,0.31]$\,mJy\,beam$^{-1}$ for the top left, top right, bottom left, and bottom right images, respectively.}
    \label{fourmfs}
\end{figure*}

The resulting MFS maps for the single ACA dataset of \cidthree (2019.2.00118.S) are shown in Figure \ref{fourmfs}.
Clearly, using only this dataset with a broad velocity range decreases the S/N dramatically.
When using different imaging parameters, the morphology changes slightly, sometimes mimicking weak extended emission. However, none of these cases show the extremely extended emission at $3\sigma$ found by \cite{cico21}, suggesting that the discrepancy is due to interactive cleaning and/or cleaning thresholds, not imaging parameters.

\subsection{Effects on combined measurement set}

\begin{figure*}
    \centering
    \includegraphics[width=0.6\textwidth]{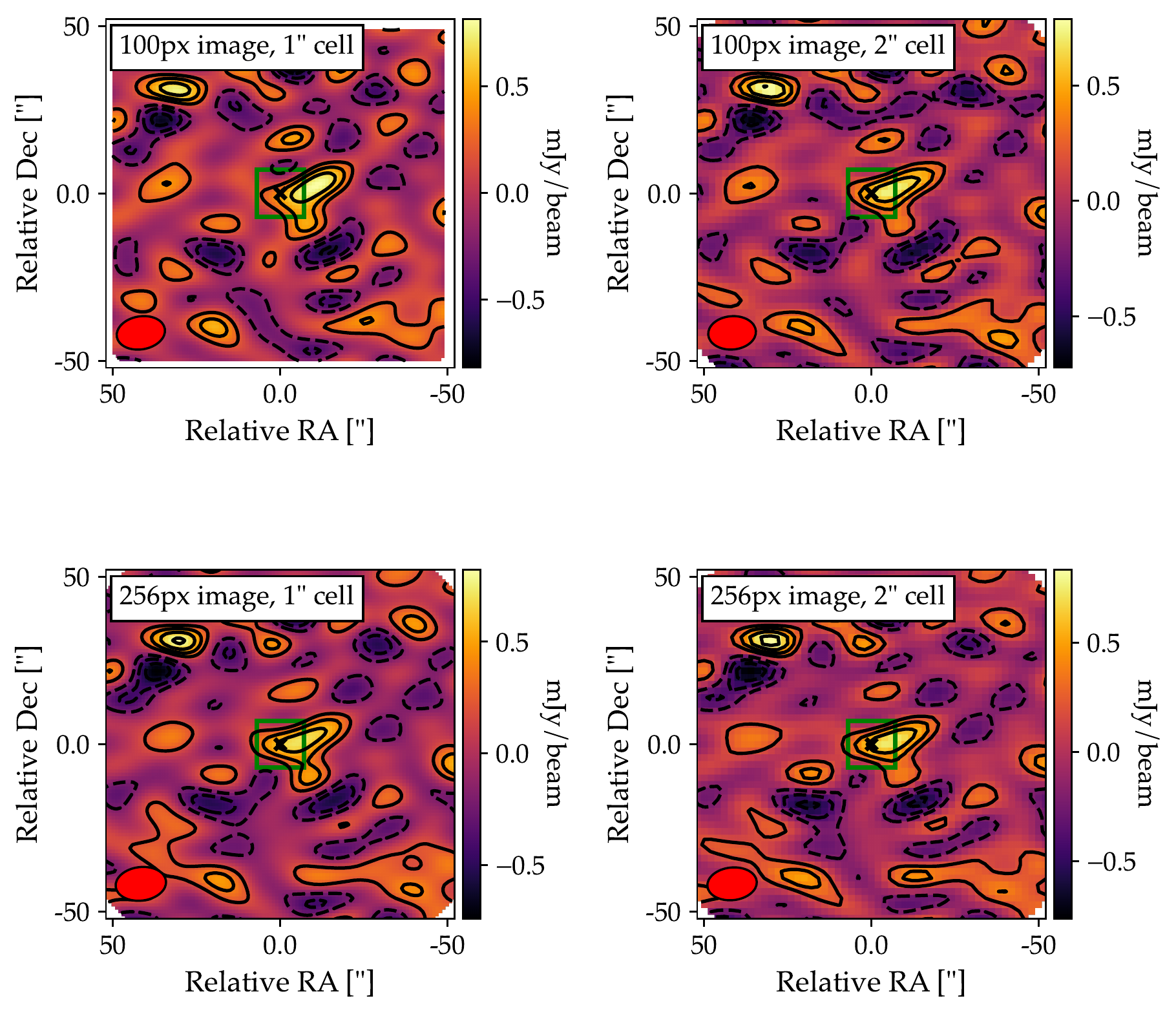}
    \caption{MFS images of CO(3-2) emission in \cidthree from ACA data. In each panel, we use the same combined dataset (2019.2.00118.S and 2021.1.00327.S) and velocity range ($-400<v<1000$\,km\,s$^{-1}$), with Briggs weighting (robust parameter of 0.5). However, we vary the cell size ($1.0''$ in the left column, $2.0''$ in the right column) and field of view ($100\times100$\,px in the top row, $256\times256$\,px in the bottom row). The contours are shown at $\pm[1,2,3,\cdots]\sigma$, where $1\sigma=[0.21,0.20,0.20,0.19]$\,mJy\,beam$^{-1}$ for the top left, top right, bottom left, and bottom right images, respectively.}
    \label{fourmfs_1}
\end{figure*}

We now use both ACA datasets of \cidthree (2019.2.00118.S and 2021.1.00327.S) to create a deeper MFS image of CO(3-2) emission using the same values of cell size and image size. The result (Figure \ref{fourmfs_1}) features a lower RMS noise level and a central $\sim3\sigma$ peak, but no evidence of extended emission on large scales. The central emission is slightly elongated to the west, but is comparable to noise peaks in the field of view, suggesting that it is not physical. Since the central detection is much weaker than that of Figure \ref{cid346spec2}, this suggests that the large velocity range dilutes the signal by including noisy channels.

\label{lastpage}
\end{document}